 \documentclass[traditabstract]{aa}
\usepackage{natbib,twoopt,amssymb,lscape}
\usepackage[breaklinks=true]{hyperref} 
\bibpunct{(}{)}{;}{a}{}{,} 
\newcommandtwoopt{\citeads}[3][][]{\href{http://adsabs.harvard.edu/abs/#3}%
{\citealp[#1][#2]{#3}}} 
\newcommandtwoopt{\citepads}[3][][]{\href{http://adsabs.harvard.edu/abs/#3}%
{\citep[#1][#2]{#3}}} 
\newcommandtwoopt{\citetads}[3][][]{\href{http://adsabs.harvard.edu/abs/#3}%
{\citet[#1][#2]{#3}}} 
\newcommandtwoopt{\citeyearads}[3][][]%
{\href{http://adsabs.harvard.edu/abs/#3}{\citeyear[#1][#2]{#3}}}
\usepackage{graphicx}
\usepackage{longtable}
\usepackage{natbib}
\bibpunct{(}{)}{;}{a}{}{,} 
\usepackage{txfonts}

\makeatletter

\newcommand{\Rmnum}[1]{\expandafter\@slowromancap\romannumeral #1@}
\makeatother
\begin{document}
   \title{Symbiotic stars in X-rays}

   \author{G. J. M. Luna,
          \inst{1}
          J. L. Sokoloski,
          \inst{2}
          K. Mukai,
          \inst{3,4}
          \and
          T. Nelson\inst{5}
          }

   \institute{Instituto de Astronom\'ia y F\'isica del Espacio (IAFE), CC 67 - Suc. 28 (C1428ZAA)  CABA -- Argentina.\\
              \email{gjmluna@iafe.uba.ar}
                  \and
           	  Columbia Astrophysics Lab, 550 W120th St., 1027 Pupin Hall, MC 5247 Columbia University, New York, New York 10027 -- USA 
\and
CRESST and X-ray Astrophysics Laboratory, NASA/GSFC, Greenbelt, MD 20771 -- USA
\and
Department of Physics, University of Maryland, Baltimore County, 1000 Hilltop Circle, Baltimore, MD 21250 -- USA
\and
School of Physics and Astronomy, University of Minnesota, 116 Church St SE, Minneapolis, MN 55409 -- USA.
}
   \date{}

  \abstract
{Until recently, symbiotic binary systems in which a white dwarf accretes from a red giant were thought to be mainly a soft X-ray population. Here we
describe the detection with the X-ray Telescope (XRT) on the {\it Swift} satellite of nine white dwarf symbiotics that were not previously known to be
X-ray sources and one that had previously been detected as a supersoft X-ray source. The nine new X-ray detections were the result of a survey of 41
symbiotic stars, and they increase the number of symbiotic stars known to be X-ray sources by approximately 30\%.  The {\it Swift}/XRT telescope detected all
of the new X-ray sources at energies greater than 2 keV.  Their X-ray spectra are consistent with thermal emission and fall naturally into three distinct
groups. The first group contains those sources with a single, highly absorbed hard component that we identify as probably coming from an accretion-disk
boundary layer. The second group is composed of those sources with a single, soft X-ray spectral component that probably originates in a region where low-velocity shocks produce X-ray emission, i.e., a colliding-wind region.

The third group consists of those sources with both hard and soft X-ray spectral components.  We also find that unlike in the optical, where rapid, stochastic brightness variations from the accretion disk typically are not seen, detectable UV flickering is a common property of symbiotic stars.  Supporting our physical interpretation of the two X-ray spectral components, simultaneous {\it Swift} UV photometry shows that symbiotic stars with harder X-ray emission tend to have stronger UV flickering, which is usually associated with accretion through a disk.  To place these new observations in the context of previous work on X-ray emission from symbiotic stars, we modified and extended the $\alpha/\beta/\gamma$ classification scheme for symbiotic-star X-ray spectra that was introduced by Muerset et al. based upon observations with the ROSAT satellite, to include a new $\delta$ classification for sources with hard X-ray emission from the innermost accretion region. Because we have identified the elusive accretion component in the emission from a sample of symbiotic stars, our results have implications for the understanding of wind-fed mass transfer in wide binaries, and the accretion rate in one class of candidate progenitors of type Ia supernovae.}

\titlerunning{Symbiotic stars}
   \maketitle
%

\section{Introduction}

Symbiotic stars are wide binary systems in which a compact object, usually a white dwarf, accretes from a more evolved companion, a red giant. Given that symbiotic stars are very heterogeneous objects, a precise observational definition naturally yields outliers \citepads[see][and references therein]{2007apn4.confE..79S}. Historically, a symbiotic system has been identified as such if its optical spectrum shows features of TiO from the red giant photosphere, emission lines of, for example, H \Rmnum{1}, He \Rmnum{2}, [O\Rmnum{3}], and sometimes a faint blue continuum. However, there are sources where some of these characteristics are not detected, either owing to the high degree of variability of symbiotics or to different system parameters such as ionizing source and nebular density. Examples of these outliers are \object{V704~Cen} where high ionization emission lines from He \Rmnum{2} or Fe \Rmnum{7} are not detected \citepads{1994A&AS..106..243C}; \object{IC~10} without emission lines from He \citepads{2008MNRAS.391L..84G}, and \object{IGR~J17197-3010} (re-identified in this work as \object{SWIFT~J171951.7-300206}, see Sect. \ref{sec:ind}) where only H lines and a red giant continuum are reported \citepads{2012A&A...538A.123M}.  

We therefore propose a physical definition where a symbiotic system is a binary in which a red giant transfers enough material to a compact companion to produce an observable signal at any wavelength. The interaction between a red giant and its compact companion can manifest itself in different ways depending upon the system parameters. Thus, our proposed definition of this class of interacting binaries is as free as possible of observational selection biases. Recognizing that a red giant can transfer material onto different types of compact companions, we refer to those that we believe have white dwarf (WD) companions as {\em WD symbiotics} and those with neutron-star (or even black-hole) companions as {\em symbiotic X-ray binaries} \citepads{2006A&A...453..295M}. In this paper, we primarily consider WD symbiotics. Because of the strong wind from the red giant, the binary system is surrounded by a dense nebula that is ionized by the UV radiation from the WD photosphere and/or the accretion disk. The orbital periods of symbiotic stars range from a few hundred to a few thousand days \citepads{2000A&AS..146..407B}. Although in WD symbiotics the white dwarfs often have masses of approximately 0.6 M$_{\odot}$ \citepads{2007BaltA..16....1M}, more massive white dwarfs, including those with masses close to the Chandrasekhar mass (M$_{Ch}$) limit, are known to exist in WD symbiotics that experience recurrent nova outbursts or produce strong, hard X-ray emission (e.g., RS~Oph, RT~Cru;
\citeads{2006Natur.442..276S, 2007ApJ...671..741L}).

The search for the progenitors of type Ia supernovae (SNIa) is currently a very active area of research. White dwarf symbiotics have been proposed as the progenitors of some SNIa either through the single or double degenerate channels. If the white dwarf in WD symbiotics can accrete at a rate that is high enough for its mass to approach $M_{Ch}$, it could become a SNIa via the single-degenerate channel (e.g., \citeads{1992ApJ...397L..87M, 2010RAA....10..235W}). \citetads{2010ApJ...719..474D} proposed that some WD symbiotics might appear as the so-called pre-double degenerate systems, before the two WDs with a total mass of greater than $M_{Ch}$ come close enough to merge within a Hubble time. There is observational evidence that at least some SNIa have a symbiotic system as a progenitor (\citeads{2007Sci...317..924P}, \citeads{2012A&A...537A.139C}, \citeads{2012Sci...337..942D}). To investigate the likelihood of WD symbiotics producing a significant fraction of SNIa, it is crucial to collect the information needed to derive basic parameters such as M$_{WD}$ and $\dot{M}$. 

Unlike cataclysmic variables (CV), where accretion is driven by Roche-lobe overflow, the accretion mechanism in symbiotics is believed to be mainly some form of wind accretion (Bondi-Hoyle; \citeads{1944MNRAS.104..273B}). Nevertheless, a consideration of angular momentum of the wind captured by the Bondi-Hoyle process leads to the conclusion that the formation of an accretion disk is common \citepads{1984Obs...104..152L, 2008ASPC..401...73W, 2011MNRAS.418.2576A}. X-ray images of the WD symbiotic $o$ Ceti show a stream of material flowing from the red giant toward the WD \citepads{2005ApJ...623L.137K}, which can be understood in the context of what is known as
the wind Roche-lobe overflow scenario proposed by \citetads{2007BaltA..16...26P}. This model suggests that even if the red giant does not fill its Roche lobe, its wind can and it is therefore focused toward the L1 point of the orbit, further increasing the likelihood of the formation of an accretion disk around the white dwarf. \citetads{2010ApJ...723.1188S}, however, found no evidence that this mechanism is enhancing the accretion rate onto the white dwarf in this binary relative to the rate expected from pure Bondi-Hoyle wind-accretion.

If an accretion disk is present, the innermost region of the accretion disk, i.e., the boundary layer, can produce X-rays. As in dwarf novae, the boundary layer in the accretion disk of WD symbiotics can be a strong source of hard ($E \gtrsim$ 2 keV) X-rays at accretion rates for which it is expected to be optically thin (e.g.,  $\dot{M} \leq$ 10$^{-9.5}$ M$_{\odot}$ yr$^{-1}$ for a 1 M$_{\odot}$ WD; \citeads{1993Natur.362..820N}). The temperature of the optically thin component, and hence the hardness of the X-ray spectrum, is expected to be a function of the gravitational potential well \citepads[see Fig. 2 in][]{2010MNRAS.408.2298B}; the more massive the white dwarf, the harder the spectrum. The combination of the mass of the accreting object and accretion rate determines what the spectrum will look like in X-rays \citepads{1982ApJS...48..239K}. White dwarfs accreting at a high rate can display a softer spectrum due probably to Compton cooling, while a harder spectrum will be detected from an equally massive white dwarf that is accreting at a lower rate. The different hardness of the X-ray spectra from RS~Oph \citepads{2011ApJ...737....7N} and T~CrB \citepads{2008ASPC..401..342L}, two recurrent novae with similar white dwarfs masses, could be explained by their different accretion rates. 

If the white dwarf magnetic field is strong, greater than a few times 10$^{5-6}$ G at the surface of the WD, hard X-rays are expected to arise from the magnetically channeled accretion flow onto a portion of the white dwarf surface. The observational signature of this type of accretion is the modulation of the light at the white dwarf spin period. In polars and intermediate polars (magnetic, accreting white dwarfs with low mass main sequence companions), the modulation is detected from optical to X-rays wavelengths \citepads{1995CAS....28.....W}. In WD symbiotics, only one system has been detected with a coherent modulation, as due to magnetic accretion, of optical emission with a period of approximately 28 m (Z~And; \citeads{2006ApJ...636.1002S}) while an oscillation with a period of 1734 s was marginally detected (95\% confidence) in X-rays from \object{R~Aqr} \citepads{2007ApJ...660..651N}.

Soft ($E \lesssim$ 2 keV), optically thin, X-ray emission in WD symbiotics can also occur in several different circumstances.  For example, soft X-rays could be produced if the system contains shocks with lower velocities than those of the shocks in a boundary layer or accretion column, as might be expected in a region where the winds from the white dwarf \citepads[e.g.,][]{2005ApJ...619..527K} or the accretion disk and red giant collide. A hot accretion disk corona, as proposed in dwarf novae \citepads[e.g.][]{2009PASJ...61S..77I}, or the red giant wind photoionized by hard X-rays would also be detected at soft X-ray energies. Although a ROSAT-based classification scheme for the X-ray spectra of symbiotic stars \citepads{1997A&A...319..201M} provided a useful framework for early work on X-ray emission from these objects, the fact that a number of symbiotics are now known to produce X-rays with energies of greater than 20 keV \citepads[e.g.,][]{2009ApJ...701.1992K,2008ApJ...675.1424C,2012arXiv1212.3336B} indicates that a new treatment of X-rays from symbiotic stars is needed. Using pointed ROSAT observations, \citetads{1997A&A...319..201M} detected 16 symbiotic stars and suggested a classification scheme based on the hardness of the spectra.  They called $\alpha$-type those systems where emission with energies of less than $\lesssim$ 0.4 keV originates in quasi-steady thermonuclear burning on the surface of the accreting white dwarf, and $\beta$-type those with X-ray spectra 
that peak at energies of about 0.8 keV that might originate in a region where the winds from the two stars collide. Because of the small bandpass of ROSAT, the X-ray spectra of sources with harder emission than the $\beta$-types were only poorly characterized; they were named $\gamma$-types. This scenario changed dramatically with the discovery of very hard X-ray emission ($E>$ 50 keV) from the symbiotic star RT~Cru with {\it INTEGRAL} \citepads{2005ATel..519....1C} and {\it Swift} \citepads{2005ATel..591....1T} in 2005. Since then, three more systems were observed to have X-ray emission with energies higher than $\approx$10 keV (T~CrB, V648~Car, CH~Cyg; \citeads{2008PASJ...60S..43S}, \citeads{2009ApJ...701.1992K}, \citeads{2007PASJ...59S.177M}). The observed spectra are all compatible with highly absorbed ($n_{H} \approx$10$^{22-23}$ cm$^{-2}$) optically thin thermal emission with plasma temperatures corresponding to $kT\approx$ 5-50 keV. Given that modulation has not been detected in their light curves, the hard X-ray emission most likely originates in the accretion disk boundary layer. The X-ray spectral fitting indicated that, like the WD symbiotics that produce softer X-rays, these hard X-ray producing symbiotics contain white-dwarf accretors. The high, variable absorption, which might be related with a clumpy medium moving into our line of sight \citepads{2009ApJ...701.1992K}, may explain why these systems were not detected in all sky surveys such as ROSAT All Sky 
Survey. In the neutron-star accretors (i.e., symbiotic X-ray binaries), the broad-band X-ray spectra are usually due to optically thick Comptonizing plasma with no emission lines \citepads[see, e.g.,][and references therein]{2011ApJ...742L..11M}.

In this article, we present the results of a {\it Swift} fill-in program whose aim was to search for hard X-ray emission from WD symbiotic, and a target of opportunity (ToO) program to identify the X-ray counterpart of \object{IGR J17197-3010}. We describe {\it Swift} observations of nine newly discovered hard X-ray emitting WD symbiotics and one previously known supersoft source.  With these new, broad-band X-ray data, it becomes necessary to introduce a classification scheme that is a modification and extension of the one proposed by \citetads{1997A&A...319..201M}. Observations and data analysis details are presented in Sect. \ref{sec:obs} while results are shown in Sects. \ref{sec:results} and \ref{sec:ind}. Section \ref{sec:disc} presents the discussion and concluding remarks.


\section{Observations and data reduction}
\label{sec:obs}

During cycle 6, {\it Swift} observed 41 symbiotics using the X-ray Telescope (XRT) and the Ultraviolet/Optical Telescope (UVOT).  We obtained these observations as part of a {\it Swift} Fill-in (6090813, PI: J. Sokoloski) and a ToO program (Target ID 31648, PI: G. J. M. Luna). Except for \object{SWIFT~J171951.7-300206}, which we found serendipitously in the field of \object{IGR~J17197-3010} \citepads{2012ATel.3960....1L}, we selected our targets from the symbiotic-star catalog of \citetads{2000A&AS..146..407B}, which lists 188 confirmed and 30 suspected symbiotics. After excluding objects with previous X-ray detections (except for StH$\alpha$ 32,  which we retained by accident), we chose the sources that are the most likely to be nearby and therefore the most likely to be detectable with {\it Swift}. Most of the objects in the \citeauthor{2000A&AS..146..407B} catalog do not have distance estimates available in the literature, so we used source brightness in the V and K bands (which are dominated by light from the red giant) as a proxy for proximity, including all objects with either V brighter than 10.9 mag (but fainter than the UVOT optical brightness limit) or K brighter than 5.0 mag. Since symbiotic stars are a disk population, objects with very large galactic latitude \textbar b\textbar\ are also preferentially nearby. Our target list thus also included all objects with \textbar b\textbar\ $> 11^{\circ}$.  {\it Swift} observed all objects for approximately 10 ks (in most 
cases using multiple visits) in photon counting mode (PC) of the XRT. The UVOT observations used either the U ($\lambda$3465 \AA, FWHM=785 \AA), UVW1 ($\lambda$2600 \AA, FWHM=693 \AA), UVM2 ($\lambda$2246 \AA, FWHM=498 \AA), and/or UVW2 ($\lambda$1938 \AA, FWHM=657 \AA) filters \citepads{2008MNRAS.383..627P}. The observation log is detailed in Table \ref{log}. In total, {\it Swift} devoted 433.6 ks to this project.

We searched for X-ray emission from each target by building images from the event files (accumulating grade 0--12 events) and using the XIMAGE package with a S/N threshold for detection of 3$\sigma$ (on average 0.0016 c s$^{-1}$ or 5.5$\times$10$^{-14}$ ergs cm$^{2}$ s$^{-1}$ assuming a thin thermal plasma with a temperature of 2 keV seeing through a 0.5$\times$10$^{22}$ cm$^{-2}$ absorption column density). All sources were detected at their catalogue positions, which were inside the {\it Swift\/}/XRT error circles (about 3 arcsec in radius). We extracted source X-ray spectra, event arrival times and light curves from a circular region with a radius of 20 pixels ($\approx$47$^{\prime\prime}$) whose centroid we determined using the tool \texttt{xrtcentroid}. To correct for the presence of dead columns on the XRT CCDs during timing analysis of XRT data, we used the standard tool \texttt{xrtlccorr}. We extracted background events from an annular region with inner and outer radii of 25 and 40 pixels, respectively. We built the ancillary matrix (ARF) using the tool \texttt{xrtmkarf} and used the \texttt{swxpc0to12s0\_20070901v011.rmf} response matrix provided by the {\it Swift} calibration team. We searched for periodic modulations in the X-ray light curves by calculating the $Z_{1}^{2}$ statistic \citepads{1983A&A...128..245B} from source event arrival times in the frequency range $f_{min}$=1/$T_{span}$ to $f_{max}$=1/(2$t_{frame}$) with a step $\Delta f$=$A/T_{span}$, where $T_
{span}$ is the difference between the last and first event arrival time (see Table \ref{log} for a list of exposure times and number of visits), $t_{frame}$ the readout time (2.5073 s for the {\it Swift}/XRT/PC), and $A$=50 the oversampling factor.

During each visit, {\it Swift} also obtained UVOT exposures in image mode. From the pipeline-reduced data, using the \texttt{uvotmaghist} script, we extracted the source count rate for each exposure from a circular region of 5$^{\prime\prime}$ radius and background from an annular region of 10$^{\prime\prime}$ and 20$^{\prime\prime}$ inner and outer radii respectively. For those objects that were not detected in individual exposures, we added the exposures using the \texttt{uvotimsum} tool to improve the detection efficiency and extracted the count rate or its upper limit using the \texttt{uvotsource} tool. No period search was performed on the UVOT data because of the small number and scarcity of the exposures on each object. We quantified the stochastic variability in the UVOT light curves by comparing the expected standard deviation from Poisson statistics only ($s_{exp}$) with the measured standard deviation ($s$) during each visit.

\begin{table*}
\caption{Observing Log. List sorted by estimated distance (see Section \ref{sec:obs}), from the nearest to the farthest away.}             
\label{log}      
\centering          
\begin{tabular}{l c c }     
\hline\hline       
Object & Exposure time [ks] & Observation Dates \\
\hline  

\object{NQ~Gem} &10.1 &2010-Apr-30  \\
\object{UV~Aur} &14.5& 2010-Apr-13/14/20 \\
\object{RW~Hya} &7.5& 2010-May-30/Aug-02\\
\object{TX~CVn} &11.5& 2010-Apr-08/21/May-05/18\\
\object{ZZ~CMi} & 11.7& 2010-May-02/04/14 \\
\object{AR~Pav} &9.6& 2010-May-20/25\\
\object{ER~Del} & 10.6& 2010-Apr-16 \\
\object{CD~-27~8661} &9.0& 2010-Aug-24/26  \\
\object{V627~Cas} &10.6& 2010-Apr-12/13  \\
\object{Hen 3-461} & 10.0& 2010-Apr-11\\
\object{WRAY~16-51} &9.3& 2010-Jul-25/Nov-11/13\\
\object{SY~Mus} &10.4& 2010-Apr-07/10/11\\
\object{CD~-283719} & 10.3& 2010-Apr-17 \\
\object{V443~Her} &9.4& 2010-May-25/Jul-01/03\\
\object{BD~-21~3873} &10.2& 2010-Aug-22/Dec-24/26/28\\
\object{NSV~05572} &10.3& 2010-Aug-01/Nov-19\\
\object{V503~Her}&9.5& 2010-Jun-04\\
\object{V748~Cen} &11.0& 2010-Sep-23, 2011-Jan-29 \\
\object{UKS~Ce-1} &11.4& 2010-Jun-26/Jul-06\\
\object{YY~Her}&9.2&2010-May-23/26/Jun-05/06\\
\object{StH$\alpha$190}&9.8&2010-Apr-03/07/13\\
\object{CI~Cyg}&18.8&2010-May-02/Jun-08/09\\
\object{FG~Ser} &9.7& 2010-Oct-07/09\\
\object{WRAY~15-1470} &10.2& 2010-Jun-26/29/Oct-13, 2011-Feb-01/02 \\
\object{Hen~3-863}&9.9&2010-Apr-13/14/15/18/21/May-04/10\\
\object{AS~210} &9.9&2011-Feb-01/02 \\
\object{StH$\alpha$ 32} & 9.9&2010-Apr-04\\
\object{V835~Cen} & 10.1&2010-Apr-19/21/May-18/21 \\
\object{BI Cru} & 10.8& 2010-Apr-11/14/15/19\\
\object{AS~289}&8.1& 2010-Jul-20/31/Aug-03\\
\object{V850~Aql} &9.8& 2010-May-30/Jun-18 \\
\object{V347~Nor} &15.7&2010-Apr-14/16/19/May-02/21\\
\object{AX~Per} &9.4& 2010-Apr-06 \\
\object{Hen~3-1213} &8.4&2010-Apr-30/May-21/26  \\
\object{LT~Del}&9.3& 2010-Jun-18/21/23/\\
\object{Y~Cra} & 9.9&2010-May-26/Jun-24 \\
\object{AS~327} &11.0&  2010-Oct-29/Nov-02/03/06 \\
\object{StH$\alpha$55} & 16.5& 2010-Apr-14/15/17/20/21\\
\object{KX~Tra}&10.6& 2010-May-22/25\\
\object{V366~Car} &8.6& 2010-Oct-08 \\
\object{SWIFT~J171951.7-300206} & 10.5 & 2012-Feb-01/02/06/08/12\\
\hline                  
\end{tabular}
\end{table*}

\section{Survey results}
\label{sec:results}

Our survey detected X-ray emission from ten sources, with spectra spanning the range of known X-ray characteristics observed in WD symbiotics.  In the ultraviolet, unlike at optical wavelengths, we detected strong flickering in most of the sources in our sample. 

\subsection{X-ray data}

The XRT detected 10 out of 41 targets in our survey, and all of the detected sources had X-ray spectra consistent with thermal emission. The soft X-ray component extends up to energies of approximately 2 keV. Detecting optically thick emission up to such energies would require very high temperatures, implying super-Eddington luminosities. For example, one of the highest temperatures detected was during the supersoft phase of the recurrent nova U~Sco, with a temperature of approximately 85 eV. The X-ray spectrum extends up to approximately 1.4 keV \citepads{2012MNRAS.tmp..361O}. We therefore conclude that the soft X-ray emission detected in the sources from our survey is due to optically thin thermal plasma. From the ten detections, one source was confirmed as a supersoft source while the remaining nine were detected at energies that exceeded 2.4 keV, with 0.3-10.0 keV count rates ranging from 0.0017 to 0.026 counts s$^{-1}$. The spectra hint at the presence of unresolved emission lines in the $\sim$1-2 keV (e.g., S \Rmnum{15}, S \Rmnum{16}, Si \Rmnum{13}, Si \Rmnum{14}, Mg \Rmnum{12}, Mg \Rmnum{11}) and $\sim$6.4 keV regions (e.g., Fe K$\alpha$, Fe \Rmnum{25}, Fe \Rmnum{26}) consistent with the presence of optically thin thermal emission. We did, however, test models of absorbed optically thick Compton plasmas \citepads[usually used to model the X-ray spectrum of symbiotic X-ray binaries; see, e.g.,][]{2007A&A...464..277M} which were discarded because the values of the 
plasma temperature were unrealistically high. 

\begin{figure*} 
\begin{center} 
\includegraphics[scale=1]{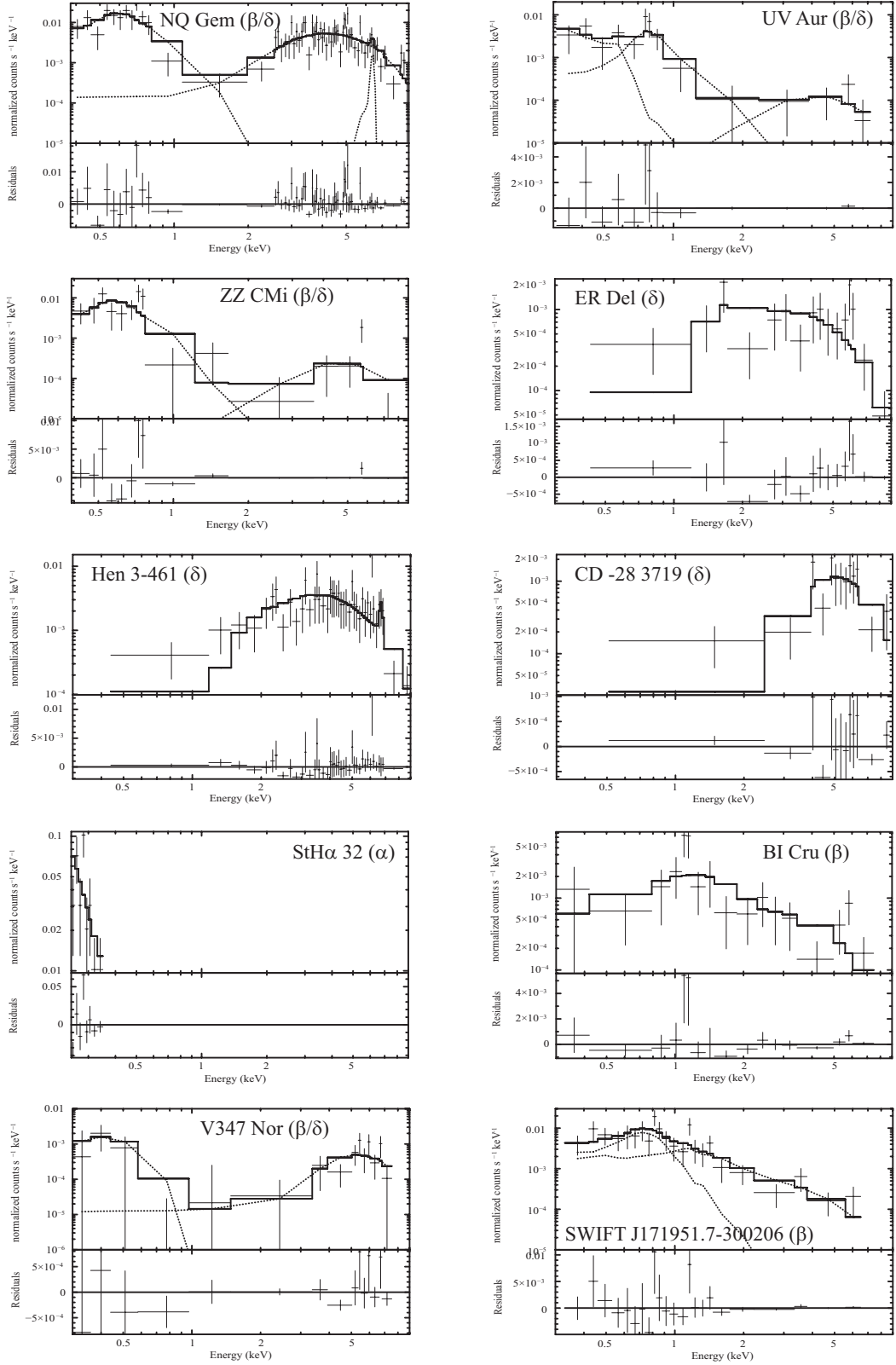} 
\caption{{\it Swift}/XRT spectra of the WD symbiotics with newly discovered X-ray emission together with their X-ray spectral types: NQ~Gem, UV~Aur, ZZ~CMi, ER~Del, Hen~3-461, CD-28~3719, StH$\alpha$ 32, BI~Cru, V347~Nor, and SWIFT~J171951.7-300206. The full line shows the best-fit model described in Section \ref{sec:results},  while the dotted line shows the contribution of the individual spectral components in the case of multi-component models. The X-ray spectral classification for each source is included between parentheses in each panel and is listed in Table \ref{tab:class}.}
\label{fig:spec} 
\end{center}
 \end{figure*}

The hardness ratios of the WD symbiotics with detectable emission above 2.4 keV ranged from $r =$ 0.14 to 9.85 (where we define $r$ as the ratio of count rates at 2.4-10.0 keV and 0.3-2.4 keV energy ranges). The X-ray spectra (Fig. \ref{fig:spec}) are consistent with optically thin thermal emission for all of the X-ray sources other than \object{StH$\alpha$~32}), with four sources showing two distinct spectral components (\object{NQ Gem}, \object{ZZ CMi},  \object{V347 Nor}, and \object{UV~Aur}) and five showing a single dominant spectral component (\object{Hen~3-461}, \object{CD-28~3719}, \object{ER~Del}, \object{BI~Cru}, and \object{SWIFT~J171951.7-300206}). Because of the low number of counts, we used the {\it C} statistic \citepads{1979ApJ...228..939C} throughout the spectral fitting procedure of the unbinned data. To determine whether the model fit the data appropriately, we calculated the goodness of fit as implemented in Xspec \citepads{1996ASPC..101...17A}, which simulates spectra many times based on the model and returns the number of simulations that have a fit statistic lower than that of the data. Ideally, if approximately 50\% of the simulations have a fit statistic lower than that from the data, then the data are well-reproduced by the model. However, some problems have been reported when fitting models with less than 100 counts in total \citepads{2011hxa..book.....A}. For those objects that we detected with less than $\approx$ 100 counts, we used visual 
inspection of the fit residuals to distinguish between two basic models, an absorbed optically thin thermal plasma or an absorbed non-thermal power law. In order to test the significance of using multi-component models (i.e., two-temperatures plasma) instead of single-component models to fit the spectrum, we followed the procedure to calculate the likelihood ratio test (LRT) described in \citetads{2002ApJ...571..545P} and implemented in XSPEC. The LRT yields the percentage of the simulations (1,000 in our case) that have a statistic (calculated as the difference between the C--stat value for the multi-component and the one for the single-component models) greater than or equal to that from fitting the data. Small values of the LRT indicate that the multi-component model is a more accurate representation of the data than the single-component model and that the presence of a second spectral component cannot arise purely  from Poisson counting statistics. Table \ref{tab:xrays} lists the resulting parameters of the spectral fitting for each object. All fit parameters are quoted at a 90\% confidence limit.  We did not detect periodic modulation in the X-ray light curves in any source of our sample. The observations were sensitive to pulsed fractions of 44\% (for \object{NQ Gem}, from which we detected the largest number of photons) or more. 

\begin{landscape} 
\begin{table} 
\caption{X-ray spectral fitting results. X-ray flux and luminosity, in units of 10$^{-13}$ ergs s$^{-1}$ cm$^{-2}$ and 10$^{31}$ ergs s$^{-1}$, respectively, are calculated in the 0.3-10.0 keV energy band. L$_{UV}$ is calculated for the bandwidth of the filter used during the observation (see Figs. \ref{fig:uvot}, \ref{fig1:uvot}, and \ref{fig2:uvot}) and tabulated in units of 10$^{31}$ ergs s$^{-1}$ (see Table \ref{time}).  }
\label{tab:xrays} 
 \begin{small}     
 \begin{tabular}{l l l l l l l l l}     
\hline\hline       
Object & Model & Count rate & $n_{H}$ & kT  & F$_{X}$ & L$_{X}$ & L$_{UV}$ & L$_{UV}/L_{X}$ \\
 & & [10$^{-2}$ counts s$^{-1}$] &[10$^{22}$ cm$^{-2}$] &  [keV] &  &   & \\
\hline  
\object{NQ~Gem} & \texttt{wabs$_{1}\times$apec$_{1}$+wabs$_{2}\times$apec$_{2}$}& 2.6$\pm$0.2 & 1: $\lesssim$0.1&1: 0.23$_{-0.03}^{+0.03}$  & 67$_{-8}^{+9}$&80$_{-9}^{+11}$ (d/1 kpc)$^2$ & 896 & $>$11 \\
 & & &2: 9.0$_{-1.7}^{+1.9}$ & 2: $\gtrsim$16& & \\
\object{UV~Aur} & \texttt{wabs$_{1}\times$(apec$_{1}$+apec$_{2}$)+wabs$_{2}\times$apec$_{3}$}& 0.29$\pm$0.04& 1: $\lesssim$0.01&1: $\lesssim$0.12; 2: 0.6$_{-0.1}^{+0.3}$ & 3.2$_{-1.7}^{+2.3}$ & 3.8$_{-2.1}^{+2.8}$ (d/1 kpc)$^2$ & $\cdots$ & $\cdots$ \\
 && & 2: 5.3$_{-3.7}^{+9.1}$& 3: $\gtrsim$2& & \\
\object{ZZ~CMi} & \texttt{wabs$_{1}\times$apec$_{1}$+wabs$_{2}\times$apec$_{2}$}&0.4$\pm$0.1 &1: $\lesssim$0.2& 1: 0.22$_{-0.05}^{+0.04}$ &6.2$_{-2.4}^{+3.1}$ &7.4$_{-2.9}^{+3.7}$ (d/1 kpc)$^2$&118&$>$16\\
 && &2: 14$_{-10}^{+19}$& $\gtrsim$2.7 & & \\
\object{ER~Del} & \texttt{wabs$\times$apec}&0.34$\pm$0.07&2$_{-1}^{+11}$ &$\gtrsim$10&5.0$_{-2.0}^{+7.0}$&6.0$_{-2.5}^{+8.0}$ (d/1 kpc)$^2$ &89 &$\gtrsim$15\\
\object{Hen 3-461} & \texttt{wabs$\times$apec}&1.4$\pm$0.1&6.1$_{-1.6}^{+2.4}$&7.6$_{-3.4}^{+11.8}$ &38$_{-9}^{+14}$ & 45$_{-9}^{+17}$ (d/1 kpc)$^2$ & 18 & $>$0.4\\
& \texttt{wabs$\times$pcfabs$\times$(mkcflow)}&&full=2.0$_{-1.3}^{+2.5}$&$\gtrsim$3.4 &58$_{-8}^{+10}$ & 70$_{-10}^{+11}$(d/1 kpc)$^2$ \\
& &&partial=10$_{-5}^{+6}$, cf=0.87$_{-0.30}^{+0.10}$&& &  \\
\object{CD~-283719} & \texttt{wabs$\times$apec}&0.30$\pm$0.05&29$_{-12}^{+20}$&$\gtrsim$11&25$_{-12}^{+19}$&30$_{-15}^{+19}$ (d/1 kpc)$^2$&167&$>$6\\
\object{StH$\alpha$ 32}\tablefootmark{a} &\texttt{bbody} &0.31$\pm$0.06&$\cdots$ & 0.03$_{-0.01}^{+0.02}$ & 8.9$_{-4.0}^{+9.0}$ & 11$_{-5}^{+11}$ (d/1 kpc)$^2$ &230&23 \\
\object{BI Cru} \tablefootmark{b} & \texttt{wabs$\times$apec}&0.4$\pm$0.1 &$\lesssim$0.3 & $\gtrsim$5 & 3.6$_{-0.9}^{+1.0}$ & 17$_{-4}^{+5}$ (d/2 kpc)$^2$& 91&5\\
\object{V347~Nor}\tablefootmark{c} &\texttt{apec$_{1}$+wabs$\times$apec$_{2}$}&0.17$\pm$0.04 &$\cdots$&0.15$_{-0.05}^{+0.06}$& 22$_{-7}^{+10}$& 59$_{-19}^{+27}$ (d/1.5 kpc)$^2$& 1080&18\\
&& &$\gtrsim$16 &$\gtrsim$2.5&& \\
\object{SWIFT~J171951.7-300206} & \texttt{wabs$\times$(apec$_{1}$+apec$_{2}$)} &0.78$\pm$0.08 &$\lesssim$0.1&1:0.3$^{+0.1}_{-0.1}$& 4.7$_{-1.6}^{+1.8}$ & 220$_{-80}^{+80}$ (d/6.3 kpc)$^2$ & $\cdots$&$\cdots$\\
& & & &2: $\gtrsim$3 & &\\
\hline                  
\end{tabular}
\tablefoottext{a}{ We used E(B-V)=0.25 \citepads{1993A&A...268..159S} to calculate the UVOT unabsorbed luminosity. } 
\tablefoottext{b}{ We used E(B-V)=1.18 \citepads{1995A&AS..111..471P} to calculate the UVOT unabsorbed luminosity. } 
\tablefoottext{c}{ We used E(B-V)=0.92 \citepads{2007apn4.confE..79S} to calculate the UVOT unabsorbed luminosity. } 

\end{small}
\end{table}
\end{landscape}

\subsection{UVOT data}

\begin{figure*}
\begin{center}
\includegraphics[scale=1.0]{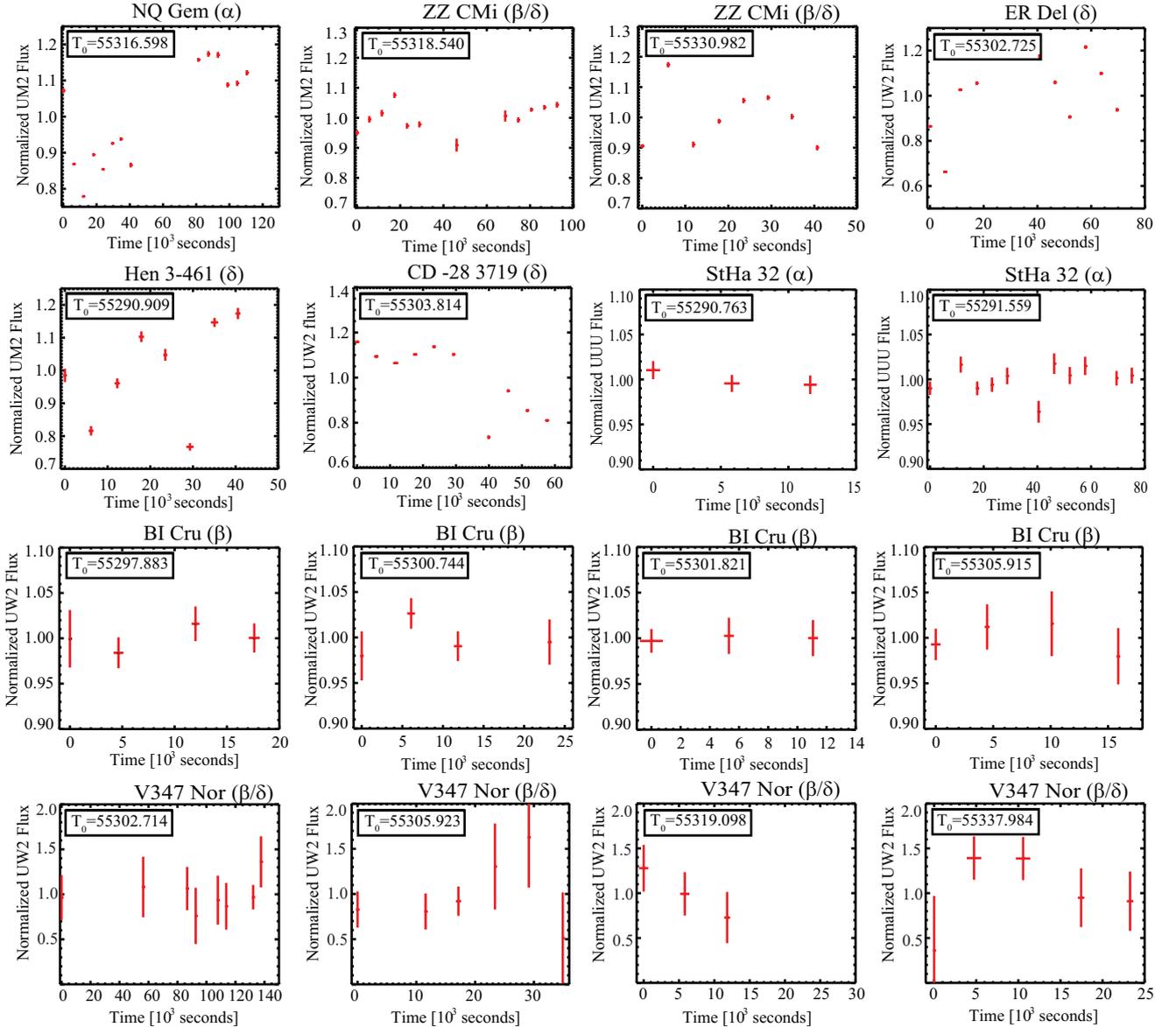}
\caption{{\it Swift} UVOT light curves of X-ray detected sources (except for UV~Aur that saturated the UVOT detector and \object{SWIFT~J171951.7-300206} that was only detectable after combining the individual exposures during each visit, see Sect. \ref{sec:results}). We show the starting time of the observation, T$_{0}$, in units of MJD. The x-axis has units of 10$^{3}$ seconds after T$_{0}$. The values in the y-axis are the fluxes normalized by the average flux of the observation. Those visits with fewer than three exposures are not shown. On each panel, we also show the X-ray spectral types (see Sect. \ref{sec:disc}) proposed for each source.}
\label{fig:uvot}
\end{center}
\end{figure*}

\begin{figure*}
 \begin{center}
 \includegraphics[scale=1.0]{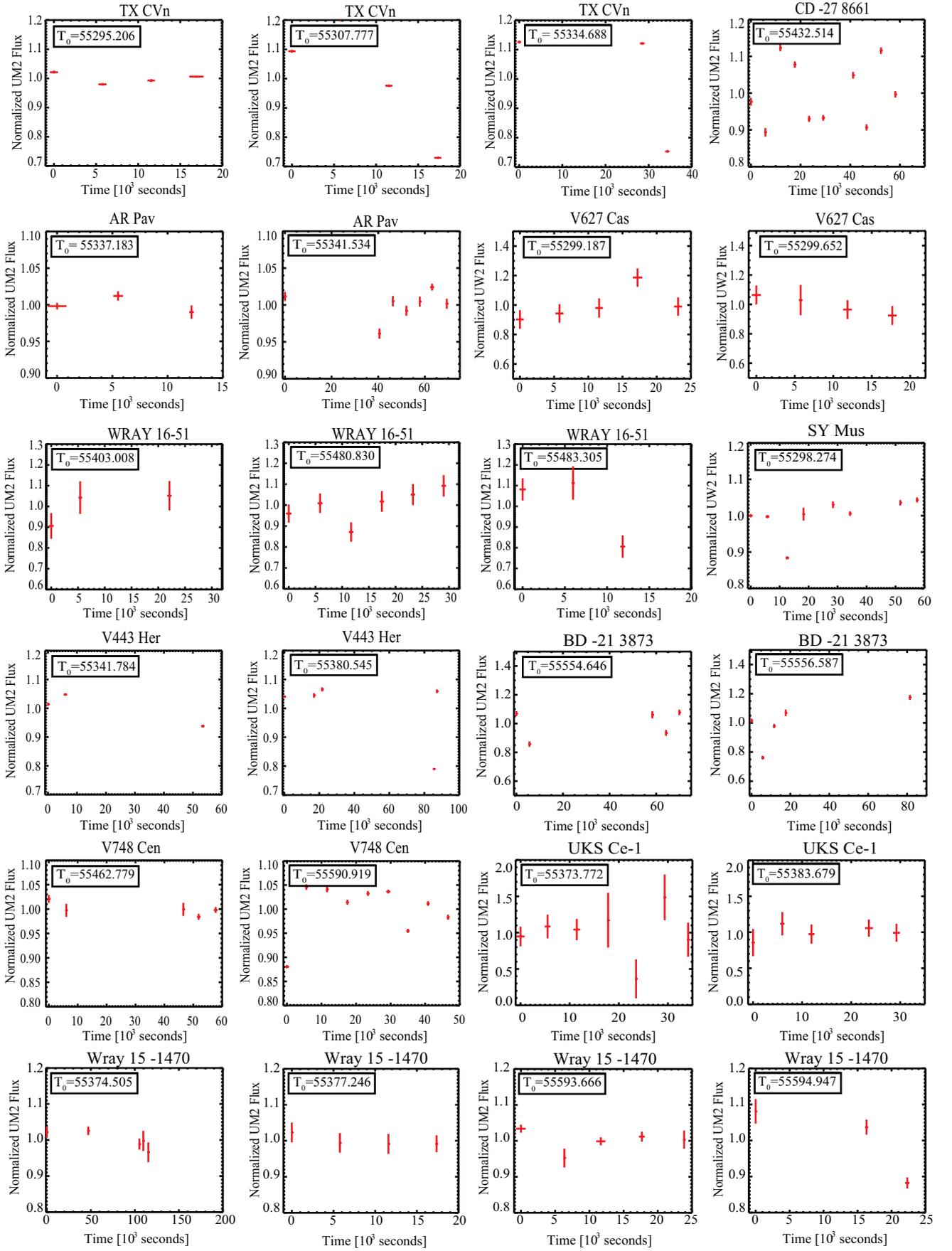}
 \caption{{\it Swift} UVOT light curves of sources that did not produce detectable X-ray emission. We show the starting time of the observation T$_{0}$ in units of MJD. The x-axis has units of 10$^{3}$ seconds after T$_{0}$. Those visits with less than three exposures are not shown. }
 \label{fig1:uvot}
 \end{center}
 \end{figure*}

\begin{figure*}
 \begin{center}
 \includegraphics[scale=1.0]{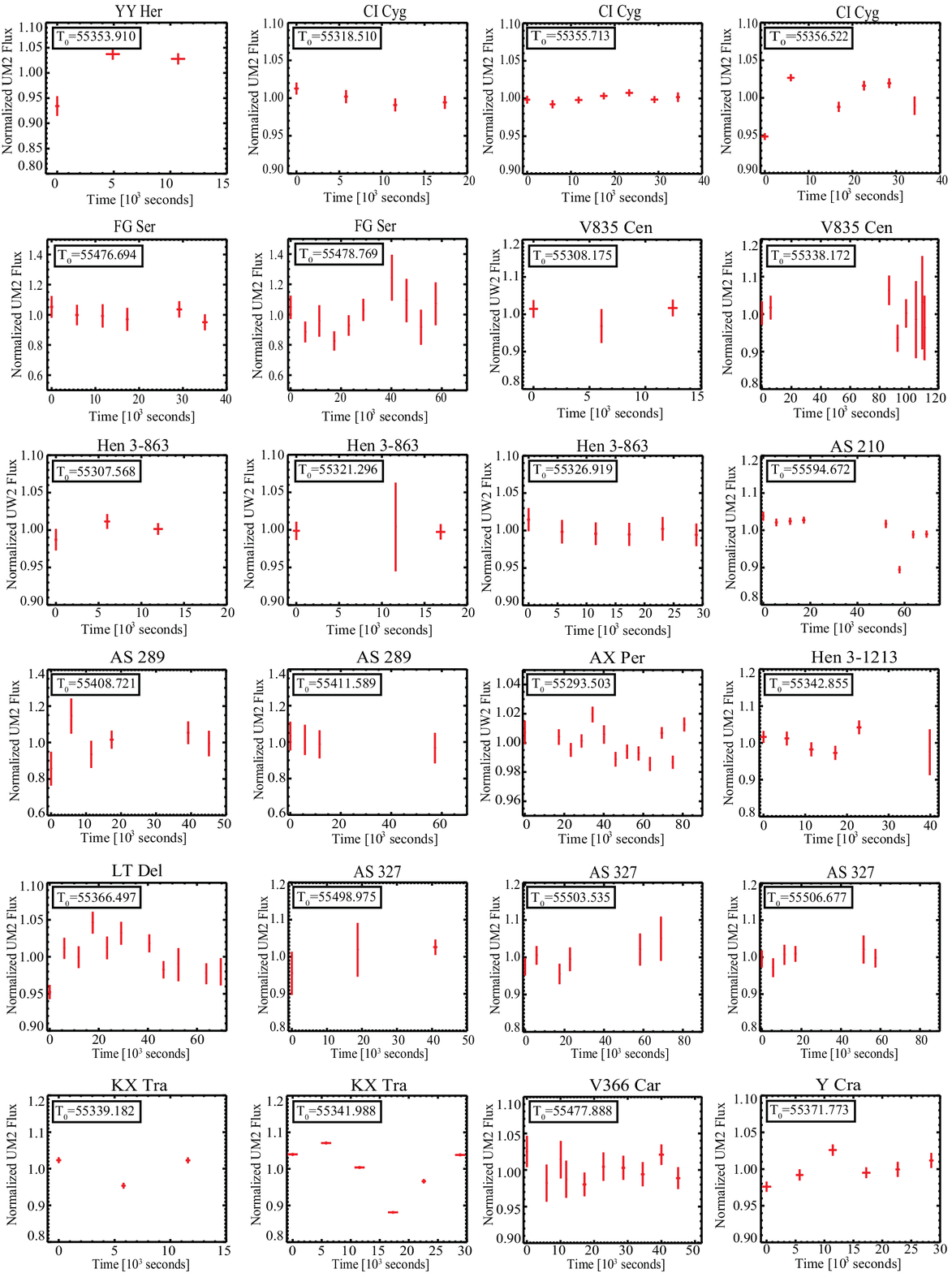}
 \caption{Same as Fig. \ref{fig1:uvot}}
 \label{fig2:uvot}
 \end{center}
 \end{figure*}

The {\it Swift} telescope detected the vast majority of our survey sources (37 out of 41) in the UV with a significance of at least 3$\sigma$. The WD symbiotics \object{UV~Aur}, \object{RW~Hya}, and \object{StH$\alpha$~190} saturated the UVOT detector so no useful UV data are available, \object{V850 Aql}, \object{V503 Her}, \object{StH$\alpha$ 55}, and \object{NSV 05572} were not detected with a 3$\sigma$ upper limit of $m_{UVM2} \gtrsim$ 21.95, \object{SWIFT~J171951.7-300206} lies inside the saturated-PSF wings of a nearby (approximately 10 arcsec away) source and was only detectable after combining the individual exposures during each visit, thus we only list the average count rate in Table \ref{time}.  Of the 33 sources with non-saturated UV detections, 21 displayed rapid variability with an rms amplitude more than twice that expected from Poisson statistics alone in at least one UV light curve segment; the rms amplitudes for these sources with unambiguous UV variability ranged from a few percent to more than 20\%. For the other 12 sources with non-saturated UV detections, the rms variability amplitude was poorly constrained in some cases (i.e., when the count rate was low), but constrained to be less than about one percent for others (see Figs. \ref{fig:uvot}, \ref{fig1:uvot}, and \ref{fig2:uvot}). Comparing the UV variability amplitude with the X-ray hardness ratio revealed that sources with the hardest X-ray spectra have the largest UV variability amplitudes (see Sect. 
\ref{sec:disc}). The observed ($s$) and expected ($s_{exp}$) standard deviations, the fractional rms variability amplitudes (on time scales of a few thousands seconds, given by the duration of each exposure), and upper limits are listed in Table 3. 

\addtocounter{table}{1}


\section{Individual objects}
\label{sec:ind}

\subsection{NQ Gem}

The star \object{NQ~Gem} is listed as a suspected symbiotic star in the catalogue of \citetads{2000A&AS..146..407B} because it shows a ratio of Si\Rmnum{3}]/C\Rmnum{3}] that is similar to that of other symbiotic stars. An orbital solution was presented by \citetads{2008AN....329...44C}, who found a period of 1308 days, an eccentricity $e$=0.182, and a lower limit on the white dwarf mass of 0.6 M$_{\odot}$. The similarity of the optical spectra of \object{NQ~Gem} and \object{T~CrB} was noted by \citetads{1971ApJ...163..309G}. 

The X-ray spectrum of \object{NQ~Gem} clearly shows two components at energies above and below $\approx$1.5 keV, respectively. This spectrum bears a striking resemblance to that of the well-known WD symbiotic CH~Cyg \citepads{2007PASJ...59S.177M}. Because of this similarity, we applied an analogous model. We fit the spectrum with a hard thermal component ($kT_{1}\gtrsim$ 16 keV) seen through a simple absorber ($n_{H,1}$=9.0$_{-1.7}^{+1.9}\times$10$^{22}$cm$^{-2}$) and the soft component with an absorbed ($n_{H,2} \lesssim$ 0.1 $\times$10$^{22}$cm$^{-2}$) low-temperature plasma ($kT_{2}$=0.23$_{-0.03}^{+0.03}$ keV). From the calculation of the LRT (see Sect. \ref{sec:obs}) we found that only 18\% of the simulations produce a statistic greater than or equal to that from fitting the data and we conclude that the two-component model is a more appropriate description of the spectrum. The unabsorbed flux is F$_{X}$[0.3-10 keV]=6.7$_{-0.8}^{+0.9}\times$10$^{-12}$ ergs cm$^{-2}$ s$^{-1}$, and the luminosity at 1 kpc (the actual distance is unknown) is L$_{X}$[0.3-10 keV]=8.0$_{-0.9}^{+1.1}\times$10$^{32}$ erg s$^{-1}$ ($d$/1 kpc)$^2$.

\subsection{UV~Aur}    

The UV Aur system is composed of UV~Aur~A and UV~Aur~B: UV~Aur~A is carbon Mira-type variable that is approximately 3.4$^{\prime\prime}$ from UV~Aur~B, a B8.5-type star \citepads{2009AJ....138.1502H}. Our UVOT observation of UV~Aur saturated the detector, which saturates for sources brighter than approximately 7.4 visual magnitudes. As UV~Aur~A has a magnitude in the range of 7.4-10.6 while UV~Aur~B has a magnitude of about 11.5 \citepads{2009AJ....138.1502H}, we conclude that the UVOT detected UV~Aur~A instead of UV~Aur~B. After the detection of [O \Rmnum{3}], [Ne \Rmnum{3}] and [Fe \Rmnum{7}] \citepads{1949PASP...61..261S,1950ApJ...111..270S, 1988syph.book..293S, 2004PASJ...56..353I}, UV~Aur~A was classified as a symbiotic star, but given the non-detection of He \Rmnum{2}$\lambda$4686 \AA~, \citetads{2009AJ....138.1502H} concluded that UV~Aur~A would not be qualified as a normal symbiotic. Nevertheless, the fact that the emission lines of [O \Rmnum{3}], etc., were detected at all is evidence that UV~Aur~A is probably a symbiotic system.

The X-ray spectrum shows two independently absorbed components, with most of the flux concentrated in the soft component. The LRT test yields values of 37\% when comparing the statistics from a single- and a two-component spectral model. We modeled the softer region of the spectrum with a weakly absorbed ($n_{H,1} \lesssim$ 10$^{20}$ cm$^{-2}$) two-temperature plasma ($kT_{1} \lesssim$0.12 keV; $kT_{2}$=0.6$_{-0.1}^{+0.3}$ keV), and the hard region was modeled with a heavily absorbed ($n_{H,2}$=5.3$_{-3.7}^{+9.1}\times$10$^{22}$cm$^{-2}$) plasma ($kT_{3} \gtrsim$ 2 keV). The unabsorbed flux is $F_{X}$=3.2$_{-1.7}^{+2.3}\times$10$^{-12}$ ergs cm$^{-2}$ s$^{-1}$. At a distance of 1 kpc \citepads{2009AJ....138.1502H}, the unabsorbed X-ray luminosity is L$_{X}$=3.8$_{-2.1}^{+2.8}\times$10$^{31}$ erg s$^{-1}$  ($d$/1 kpc)$^2$.

\subsection{ZZ CMi}

Although \citeauthor{2000A&AS..146..407B} were ambivalent about whether ZZ~CMi is a symbiotic star, the similarities between the properties of the X-ray and UV emission from ZZ~CMi and those of other well-established WD symbiotics leads us to conclude that it is indeed a WD symbiotic.  \citetads{2000A&AS..146..407B} noted that the optical colors do not evolve like those of other symbiotics, that the optical emission line strengths are unusual (H$\gamma$ $>$ H$\beta$), and that the maximum ionization potential could be as low as 35.1 eV, but the source definitely contains a late-type star and displays an emission-line optical spectrum, and the H$\alpha$ profile is similar to that of other WD symbiotics.  Since some WD symbiotics with very hard X-ray spectra can have optical spectra that appear to be only weakly symbiotic, ZZ~CMi could provide another example of the different views of WD symbiotics provided by X-ray and optical observations.

As in the case of \object{NQ Gem}, the X-ray spectrum from \object{ZZ CMi} closely resembles the spectrum from \object{CH~Cyg}, with two components primarily above and below $\approx$2 keV. The LRT indicates that only in 21\% of the simulations was a simpler model acceptable over the more complex model that we used. Therefore we applied a similar spectral model to the one used for NQ~Gem consisting of a weakly absorbed ($n_{H,1}\lesssim$ 0.2$\times$10$^{22}$ cm$^{-2}$) optically thin thermal plasma ($kT_{1}$=0.22$_{-0.05}^{+0.04}$ keV) to model the softer energies plus an absorbed ($n_{H,2}=$14$^{+19}_{-10}\times$10$^{22}$cm$^{-2}$) optically thin plasma ($kT_{2}\gtrsim2.7$ keV) at higher energies. The unabsorbed flux is F$_{X}$=6.2$_{-2.4}^{+3.1}\times$10$^{-13}$ ergs cm$^{-2}$ s$^{-1}$, and the luminosity at 1 kpc is L$_{X}$=7.4$_{-2.9}^{+3.7}\times$10$^{31}$ ergs s$^{-1}$ ($d$/ 1 kpc)$^2$.

\subsection{ER Del}

Although the spectral type of the cool component in ER~Del is S5.5/2.5 \citepads{1979ApJ...234..538A}, which is relatively rare for a symbiotic star \citepads{2002A&A...396..599V}, the optical and UV emission lines \citepads{2000A&AS..146..407B} support a symbiotic-star classification. In symbiotic stars that contain S stars, the ZrO bands in the spectrum of the red giant indicate that the red giant has been polluted by mass transfer from the companion \citepads{1999A&A...345..127V}. The UV emission lines have ionization potentials as high as 47.9 eV, and the optical spectrum shows emission lines of H \citepads{2000A&AS..146..407B}. Moreover, \citetads{2012BaltA..21...39J} recently determined an orbital period for ER~Del of $2089\pm6$~d. These features would suggest that ER~Del is indeed a symbiotic binary.
 
The small number of photons detected (36 photons in a 10.6 ks exposure time) did not allow us to perform a precise fit. We applied a simple model consisting of an absorbed ($n_{H}=2^{+11}_{-1}\times$10$^{22}$cm$^{-2}$) optically thin thermal plasma ($kT \gtrsim$ 10 keV). The unabsorbed flux is F$_{X}$=5.0$_{-2.0}^{+7.0}\times$10$^{-13}$ ergs cm$^{-2}$ s$^{-1}$, and the luminosity at 1 kpc is L$_{X}$=6.0$_{-2.5}^{+8.0}\times$10$^{31}$ ergs s$^{-1}$ ($d$/1 kpc)$^2$. The large value of the absorbing column is indicated by the low count rate below $\leq$ 2 keV.

\subsection{Hen 3-461}
\label{sec:hen}


The star \object{Hen 3-461} was classified as a suspected symbiotic in the catalog of \citetads{1984PASAu...5..369A}. Its optical spectrum shows a late-type continuum with prominent TiO bands and emission lines from the Balmer series, He \Rmnum{1}, [Ne \Rmnum{3}], [O \Rmnum{3}], and [Fe\Rmnum{7}]. The optical spectrum of \object{Hen 3-461} resembles the spectrum of RT~Cru and T~CrB in quiescence \citepads{1998AJ....116.1971P}, with a strong red continuum and weak Balmer lines. Little is known about this source at other wavelengths.

The X-ray spectrum of \object{Hen 3-461} (Fig. \ref{fig:spec}) is similar to the spectrum of RT~Cru \citepads{2007ApJ...671..741L,2009ApJ...701.1992K} in that it consists of a highly absorbed, strong continuum extending to high energies. Assuming a simple model consisting of an absorbed, optically thin thermal plasma, we find $n_{H}$=6.1$_{-1.6}^{+2.4}\times 10^{22}$ cm$^{-2}$ and $kT$=7.6$_{-3.4}^{+11.8}$ keV. This model has a unabsorbed flux F$_{X}$=3.8$_{-0.9}^{+1.4}\times$10$^{-12}$ ergs cm$^{-2}$ s$^{-1}$. Taking a more complex model, similar to \object{RT Cru} and \object{T CrB} \citepads{2007ApJ...671..741L,2008ASPC..401..342L},  which consists of an absorbed multi-temperature cooling flow plasma, we find a lower limit for the maximum temperature for the cooling flow component of $kT_{max} \gtrsim$ 3.4 keV and solar abundances \citepads{1989GeCoA..53..197A}. In the complex model, the absorber has two components, one that completely covers the X-ray source ($n_{H}$(full)=2.0$_{-1.3}^{+2.5}\times 10^{22}$ cm$^{-2}$) and one that only partially covers it ($n_{H}$(partial)=10$_{-5}^{+6} \times 10^{22}$ cm$^{-2}$, with a covering fraction of 0.87$_{-0.30}^{+0.10}$). Assuming a distance of 1 kpc, the resulting mass accretion rate is $\dot{M}\lesssim 4\times10^{-9}$ M$_{\odot}$/yr ($d$/1 kpc)$^{2}$. The measured unabsorbed flux is F$_{X}$=5.8$_{-0.8}^{+1.0}\times$10$^{-12}$ ergs cm$^{-2}$ s$^{-1}$ and so the luminosity is L$_{X}$=7.0$_{-1.0}^{+1.1}\times$10$^{32}$ ergs s$^{-
1}$ ($d$/1 kpc)$^2$. The difference in the flux from the one-temperature model and the cooling flow can be attributed to the difference in the amount of absorption.

\subsection{CD -28 3719}

The symbiotic nature of \object{CD~-28 3719} has been suggested based on its broad H$\alpha$ profiles and blue colors (\citeads{2000A&AS..146..407B} and references therein). With an exposure time of 10.2 ks, we detected 30 X-ray photons from \object{CD~-28 3719}. We fit the spectrum with a simple model composed of a highly absorbed (n$_{H}=29_{-12}^{+20}\times$10$^{22}$ cm$^{-2}$) plasma with a temperature of $kT \gtrsim$ 11 keV. The unabsorbed flux is F$_{X}$=2.5$_{-1.2}^{+1.9}\times$10$^{-12}$ ergs cm$^{-2}$ s$^{-1}$, and the luminosity at a distance of 1 kpc is L$_{X}$=3.0$_{-1.5}^{+1.9}\times$10$^{32}$ ergs s$^{-1}$ ($d$/1 kpc)$^2$. Although the low number of photons precludes a more precise fit, the lower limit on $n_{H}$ requires the spectrum to be highly absorbed. 

\subsection{StH$\alpha$ 32}

The star StH$\alpha$ 32 is a known supersoft source \citepads{1996LNP...472..225B,2007ApJ...661.1105O} and it was included by accident in our target list. However, no X-ray spectrum has been published in the literature until now. Based on a method to determine the probability of 2MASS (Two Micron All Sky Survey) and ROSAT sources being associated, \citetads{2009ApJS..184..138H} determined a probability of 0.721 that StH$\alpha$ 32 is associated with the symbiotic source \object{2MASS J0437456-0119118} (see Haakonsen \& Rutledge 2009 for details about the method used). The system \object{StH$\alpha$ 32} belongs to the small group of barium-rich symbiotics, i.e., systems that exhibit symbiotic features such as H \Rmnum{1} and He \Rmnum{2} in their optical and UV spectra and barium-star-type abundance anomalies \citepads{1994A&A...284..156S}. 

Given that {\it Swift}/XRT detected only 31 photons from \object{StH$\alpha$ 32}, all with energies less than or equal to 0.4 keV, we only obtained approximated values for the parameters of the spectral model. We fit the spectrum with a blackbody model with a temperature of $kT$=0.03$_{-0.01}^{+0.02}$ keV (absorption was negligible). The flux is F$_{X}$=8.9$_{-4.0}^{+9.0}\times$10$^{-13}$ ergs cm$^{-2}$ s$^{-1}$ and at a distance of 1 kpc, the luminosity is L$_{X}$=1.1$^{+1.1}_{-0.5}\times$10$^{32}$ ergs s$^{-1}$ ($d$/1 kpc)$^2$. Most known supersoft sources have luminosities in the 10$^{35-36}$ ergs s$^{-1}$ range \citepads[see, e.g.,][]{2007ApJ...661.1105O}; therefore, it is possible that StH$\alpha$ 32 is farther away, probably in the galactic halo, as proposed by \citetads{1993A&A...268..159S}, based on the small reddening toward the source, its galactic coordinates ($l$ = 197$^{\circ}$, $b$ = -30$^{\circ}$), and radial velocity ($v_{r}$=325 km s$^{-1}$).

\subsection{BI~Cru}

The symbiotic system \object{BI~Cru} is comprised of a Mira-type red giant with a pulsation period of 280 days, an accreting white dwarf, and a bipolar nebula that extends 1.3 pc from the central binary perpendicular to the orbital plane \citepads{2009MNRAS.396..807C}. The bipolar structures (expanding at $\approx$200 km s$^{-1}$) could be explained by the presence of an accretion disk and periodic hydrogen shell flashes on the surface of the white dwarf (with flashes every $\sim$ 1000 yr; \citeads{1993A&A...268..714C}). In their model of an optical spectrum taken in 1974, \citetads{2009MNRAS.396..807C} proposed that shocks in the inner nebula (from an unrecorded outburst) could be fast enough, with speeds of a few thousands km s$^{-1}$, to produce X-ray emission. If these shocks produced the X-ray emission that we observed, they either must not have had time to cool or must have been fed by more recent mass ejections.

We fit the X-ray spectrum with a simple model of an absorbed ($n_{H}\lesssim 0.3\times 10^{22}$ cm$^{-2}$) optically thin thermal plasma ($kT \gtrsim$ 5 keV) with non solar ($\gtrsim$ 2) Ne abundance. The total unabsorbed flux is F$_{X}$=3.6$_{-0.9}^{+1.0}\times$10$^{-13}$ ergs cm$^{-2}$ s$^{-1}$ and at a distance of 2 kpc \citepads{2008ApJ...682.1087M} the luminosity is L$_{X}$=1.7$_{-0.4}^{+0.5}\times$10$^{32}$ ergs s$^{-1}$ ($d$/ 2 kpc)$^2$. The residuals at energies of about 1 and 6 keV (see Fig. \ref{fig:spec}) suggest that the X-ray emission could arise from a multi-temperature plasma, however based on the LRT test, there is not significant improvements in the statistic when using multi-component spectral models. 

\subsection{V347~Nor}

The star \object{V347~Nor} is a symbiotic with a Mira-type red giant \citepads{2000A&AS..146..407B}. It shows an extended nebula discovered by \citetads{1993A&A...277..195M}. \citetads{2007A&A...465..481S} determined a distance of 1.5$\pm$0.4 kpc using the expansion parallax method.  Based on the similarity of the X-ray spectrum with \object{CH~Cyg}, we fit the X-ray spectrum with a two-component model. The LRT test indicates that only 13\% of the simulations yielded a statistic equal to or smaller than the statistic from a two-component model. We used two optically thin thermal plasmas: a low temperature plasma ($kT_{1}$=0.15$_{-0.05}^{+0.06}$ keV) and a highly absorbed ($n_{H}\gtrsim$16$\times$10$^{22}$ cm$^{-2}$) high temperature plasma ($kT_{2} \gtrsim$ 2.5 keV). The unabsorbed flux is F$_{X}$=2.2$_{-0.7}^{+1.0}\times$10$^{-12}$ ergs cm$^{-2}$ s$^{-1}$, and the luminosity at 1.5 kpc is L$_{X}$=5.9$_{-1.9}^{+2.7}\times$10$^{32}$ ergs s$^{-1}$ ($d$/ 1.5 kpc)$^2$.

\subsection{SWIFT~J171951.7-300206, a newly discovered symbiotic in the field of IGR~J17197-3010}

In February 2012, {\it Swift}/XRT detected an X-ray source at the coordinates $\alpha$ = 17h 19m 51.7s and $\delta$=-30$^{\circ}$ 02$^{\prime}$ 0.6$^{\prime\prime}$ \citepads[with an error radius of 4.3$^{\prime\prime}$,][]{2012ATel.3960....1L}. These XRT coordinates are consistent with the position of a symbiotic star at $\alpha$=17h 19m 51.83s and $\delta$=-30$^{\circ}$ 02$^{\prime}$ 0.3$^{\prime\prime}$ \citepads{2012A&A...538A.123M}. We therefore use the {\it Swift} naming convention and hereafter refer to this symbiotic as \object{SWIFT~J171951.7-300206}. Although \citetads{2012A&A...538A.123M} proposed that this symbiotic star might be the counterpart to the $\gamma$-ray source IGR~J17197-3010, \citetads{2012ATel.3960....1L} concluded that the location of the two X-ray sources in the {\it Swift}/XRT field of the $\gamma$-ray source did not support the association between the symbiotic star and the $\gamma$-ray source. Therefore, although WD symbiotics have been known to produce $\gamma$-rays \citepads[e.g.,][]{2005ATel..528....1M}, \object{SWIFT~J171951.7-300206} appears unlikely to have done so.

The XRT spectrum of \object{SWIFT~J171951.7-300206} extends up to approximately 5 keV. We model the spectrum with an absorbed ($n_{H} \lesssim$ 0.1$\times$10$^{22}$ cm$^{-2}$) two-temperature plasma ($kT^{1}$=0.3$_{-0.1}^{+0.1}$ keV and $kT^{2} \gtrsim$3 keV). The LRT test indicates that in 33\% of the simulations the statistic of a single-component model was equal to or smaller than the statistic of a two-component model. The unabsorbed flux is F$_{X}$=4.7$_{-1.6}^{+1.8}\times$10$^{-13}$ ergs cm$^{-2}$ s$^{-1}$, and at a distance of 6.3 kpc \citepads{2012A&A...538A.123M}, the X-ray luminosity is L$_{X}$=2.2$_{-0.8}^{+0.8}\times$10$^{33}$ ergs s$^{-1}$ ($d$/ 6.3 kpc)$^2$.

\section{Discussion and conclusions}
\label{sec:disc}

We find that the X-ray spectra of newly discovered X-ray sources fall naturally into three groups.  The first comprises those sources with highly absorbed, hard ($E \gtrsim$ 2 keV) single-component X-ray spectra.  The second includes sources with two distinct X-ray spectral components, one soft ($E \lesssim$ 2 keV) and one hard.  The third group is made up of sources with soft, single-component X-ray spectra. As the $\alpha$, $\beta$, and $\gamma$ categorization introduced by \citetads{1997A&A...319..201M} was based on ROSAT data, it naturally missed those WD symbiotics with hard, highly absorbed X-ray spectra. Moreover, the hard component of those systems with both soft and hard X-ray spectral components were also not detectable with ROSAT, and two-component X-ray spectra were thus also not included in this scheme.

We therefore propose an updated classification scheme for the X-ray spectra of symbiotic stars that builds upon and extends the previous scheme proposed by \citetads{1997A&A...319..201M}. We retain their $\alpha$, $\beta$, and $\gamma$ X-ray spectral classes and introduce a new category, that we have called $\delta$, to identify those WD symbiotics with hard, highly absorbed X-ray spectra. Since WD symbiotics with both soft and hard components in their X-ray spectra share features of the $\beta$- and $\delta$-types, we dub these systems $\beta/\delta$. We summarize the groups as:

\begin{description}
\item[$\alpha$:] Supersoft X-ray sources with most of the photons having energy less than 0.4 keV (all photons are detected below 1 keV). The likely origin is quasi-steady shell burning on the surface of the white dwarf \citepads[e.g.,][]{2007ApJ...661.1105O}.
\item[$\beta$:] Soft X-ray sources with most of the photons having energy less than 2.4 keV, the maximum energy detectable with ROSAT. The likely origin is the collision of winds from the white dwarf with those from the red giant \citepads{1997A&A...319..201M}. 
\item[$\gamma$:] Symbiotic stars with neutron-star accretors, also known as symbiotic X-ray binaries. Their X-ray spectra extend toward high energies (E $\gtrsim$ 2.4 keV) and can be modeled as due to optically thick Comptonized plasma \citepads[e.g.,][]{2007A&A...470..331M}. 
\item[$\delta$:] Highly absorbed, hard X-ray sources with detectable thermal emission above 2.4 keV. The likely origin is the boundary layer between an accretion disk and the white dwarf. 
\item[$\beta/\delta$:] WD symbiotics with two X-ray thermal components, soft and hard. They share features of $\beta$ and $\delta$ types. The soft emission is most likely produced in a colliding-wind region \citepads{1997A&A...319..201M}, and the hard emission is most likely produced in an accretion-disk boundary layer (see Sect. 5.2).
\end{description}

\begin{figure}
\includegraphics[scale=0.53]{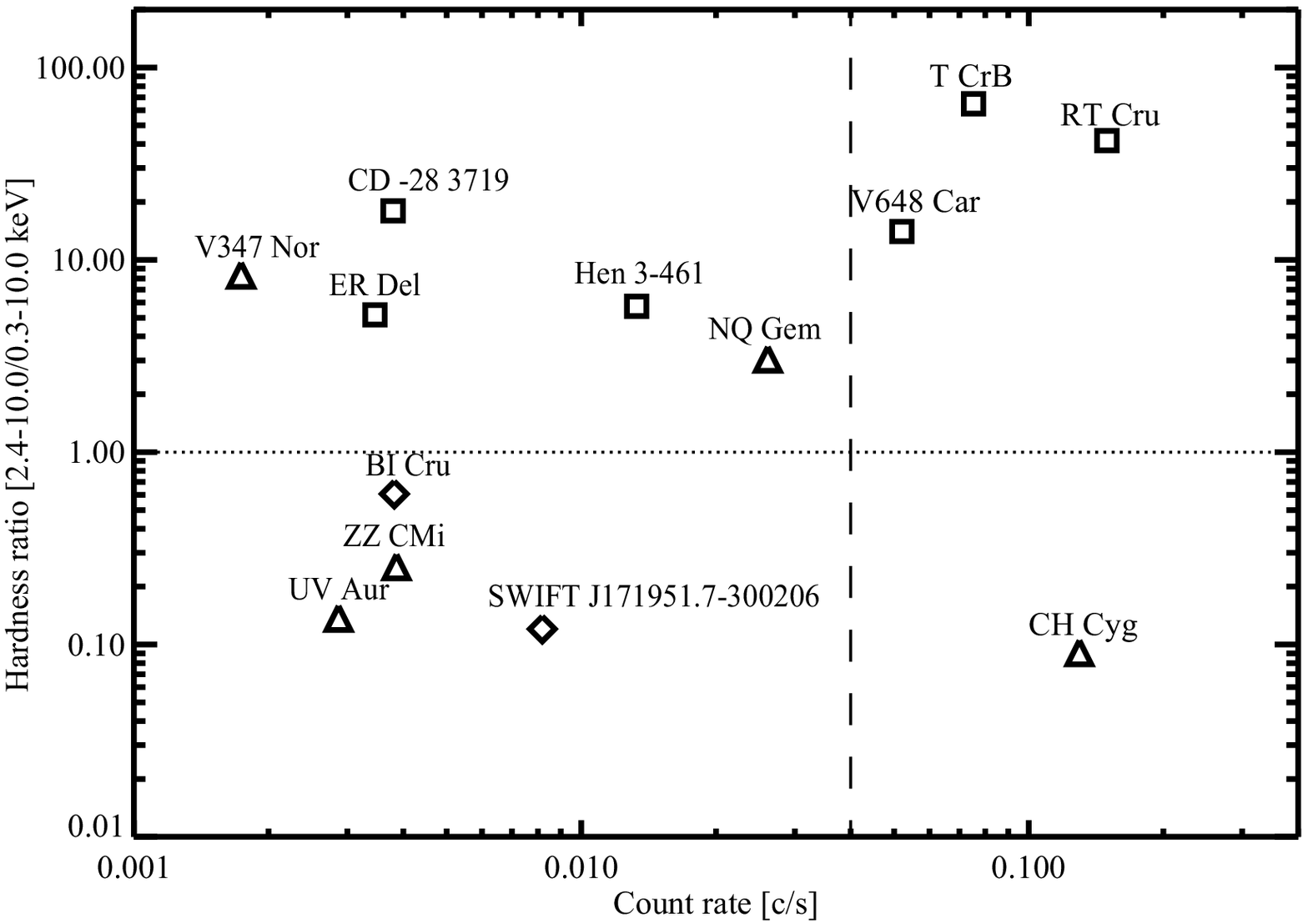}
\caption{Hardness ratio vs count rate in counts s$^{-1}$ for the hard X-ray WD symbiotics that have been observed with {\it Swift}. The new X-ray detected WD symbiotics are all located to the left of the dashed line, confirming that because of their low X-ray fluxes, they had not been detected before. The plot shows that $\delta$-type sources ($\square$) have a {\em hardness ratio} of more than 1 (above the dotted horizontal line); $\beta$-type sources ($\diamond$) have {\em hardness ratio} of less than 1; and $\beta/\delta$-type sources ($\triangle$) are located above and below the {\em hardness ratio} $=$ 1 line. Hence, accretion-dominated $\delta$-type objects lie above {\it hardness ratio} $\approx$ 1, while below this line we find soft X-ray sources whose X-ray spectra are dominated by emission originated in a colliding-wind region.}
\label{fig:hr}
\end{figure}

In Table \ref{tab:class} we show the classification, under the new scheme, of all the symbiotics that have reported X-ray detections. Some classifications are uncertain owing to the short exposure time of our exploratory survey and are labeled as such in Table \ref{tab:class}. Figure \ref{fig:hr} shows X-ray hardness (as defined in Sect. \ref{sec:results}) as a function of XRT count rate for the WD symbiotics with newly detected X-ray emission as well as those with previously known $\delta$-type emission \citepads[RT~Cru, T~CrB, V648~Car, and CH~Cyg;][]{2009ApJ...701.1992K}. As expected, the sources with newly detected X-ray emission have lower fluxes than the prior discoveries, confirming that they had not been detected in various previous X-ray surveys because they were too faint. We can also see in this figure that there are basically two regions, above and below {\it hardness ratio} $\approx$ 1, that separate $\delta$-type objects from $\beta$-type objects.

\begin{table*}[h]
\caption[]{X-ray spectral classifications of symbiotic stars.}
\begin{scriptsize}
\begin{tabular}{l c c }
\hline\hline
Object &  Type & Reference \\
\hline
\hline
\hline
\object{StH$\alpha$ 32} &$\alpha$ & 1, this work\\
 \object{SMC~3} & $\alpha$  & 2\\
 \object{Ln~358} & $\alpha$  & 2\\
 \object{AG~Dra} & $\alpha$  & 2\\
 \object{Draco~C-1} & $\alpha$  & 2\\
 \object{RR~Tel} & $\alpha$  & 2\\
 \object{CD-43~14304} & $\alpha$  & 2\\
\object{BI Cru}\tablefootmark{a} & $\beta/\delta$ & this work\\
\object{SWIFT~J171951.7-300206} & $\beta$& this work\\
 \object{RX~Pup}& $\beta$ & 2, 3  \\
 \object{Z~And} & $\beta$ & 2, 4\\
 \object{V1329~Cyg} & $\beta$  & 5\\
 \object{Mira~AB}& $\beta$ & 6\\
 \object{EG~And} & $\beta$  & 2\\
 \object{HM~Sge} & $\beta$  & 2\\
 \object{V1016~Cyg} & $\beta$  & 2\\
 \object{PU~Vul} & $\beta$  & 2\\
 \object{AG~Peg} & $\beta$  & 2\\
 \object{Hen~2-104} & $\beta$ & 21\\
 \object{Hen~3-1341} & $\beta$ & 22\\
\object{NQ~Gem} & $\beta/\delta$ & this work\\
\object{UV~Aur} & $\beta/\delta$ & this work\\
\object{ZZ~CMi} & $\beta/\delta$ & this work\\
\object{V347~Nor} &$\beta/\delta$& this work\\
 \object{R~Aqr}& $\beta/\delta$ & 2, 7 \\
 \object{CH~Cyg}&  $\beta/\delta$ & 2, 8 \\
 \object{MWC~560}& $\beta/\delta$ & 20 \\
\object{ER~Del} & $\delta$  & this work\\
\object{Hen 3-461} & $\delta$ & this work\\
\object{CD~-283719} & $\delta$ & this work\\
 \object{RT~Cru} & $\delta$ & 9, 10\\
 \object{T~CrB}& $\delta$ & 11 \\
 \object{V648~Car}& $\delta$ & 12, 13 \\
 \object{GX~1+4} & $\gamma$  & 2\\
 \object{Hen~3-1591}\tablefootmark{b} & $\gamma$  & 2\\
\object{V934~Her} & $\gamma$ & 14\\
 \object{4U~1954+31} & $\gamma$ & 15\\
 \object{Sct~X-1} & $\gamma$ & 16\\
 \object{IGR~J16194-2810} & $\gamma$ & 17\\
 \object{IGR~J16358-4726}\tablefootmark{c} & $\gamma$ & 19 \\
 \object{IGR~J16393-4643} & $\gamma$ & 18 \\
 \object{CGCS~5926} & $\gamma$ &  19\\
\label{tab:class}
\end{tabular}
\tablebib{(1)~\citetads{2007ApJ...661.1105O};~(2)\citetads{1997A&A...319..201M}; (3)~\citetads{2006Ap&SS.304..283L};(4)~\citetads{2006ApJ...636.1002S}; (5)~\citetads{2011ApJ...731...12S};(6)~\citetads{2010ApJ...723.1188S}; (7)~\citetads{2007ApJ...660..651N}; (8)~\citetads{2007PASJ...59S.177M}; (9)~\citetads{2007ApJ...671..741L}; (10)~\citetads{2009ApJ...701.1992K}; (11)~\citetads{2008ASPC..401..342L};(12)~\citetads{2010ApJ...709..816E}; (13)~\citetads{2008PASJ...60S..43S}; (14)~\citetads{2002A&A...382..104M};( 15)~\citetads{2006A&A...453..295M}; (16)~\citetads{2007ApJ...661..437K}; (17)~\citetads{2007A&A...470..331M}; (18)~\citetads{2006ApJ...649..373T}; (19)~\citetads{2011A&A...534A..89M}; (20)~\citetads{2009A&A...498..209S}; (21)~\citetads{2006AAS...209.9206M}; (22)~\citetads{2013A&A...554A..56S}}

\tablefoottext{a}{Questionable classification due to short exposure time. There are hints of the presence of a second soft spectral component, but it needs confirmation.}
\tablefoottext{b}{Questionable classification.  Hen~3-1591 has been observed only with ROSAT, therefore no information is available about its hard X-ray emission, and the nature of the accreting object is not firm enough to secure its classification.}
\tablefoottext{c}{Questionable classification. \citetads{2008A&A...484..783C} suggest that \object{IGR~J16358-4724} is a high-mass X-ray binary, however \citetads{2010A&A...516A..94N} suggest a symbiotic nature.}
 \end{scriptsize}
\end{table*}

\begin{figure}
\includegraphics[scale=0.53]{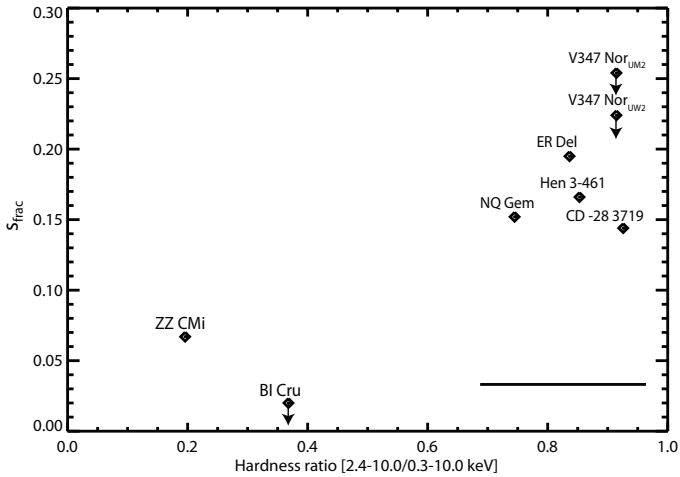}
\caption{Fractional rms amplitude of rapid UV variability ($s_{frac}$) vs ratio of hard (2.4-10.0 keV) to total (0.3-10.0 keV) X-ray count rates. Objects with harder X-ray spectra tend to have more intense UV variability. Since rapid variability is a hallmark of accretion, this trend supports our proposition that the hard X-ray emission in WD symbiotics is powered by accretion. \object{V347~Nor} was observed with two UV filters, and we plot the fractional variability from each of the observations. Downward arrows indicate upper limits. The average error bar is shown at the bottom-right corner.}
\label{fig:ssexp_hr}
\end{figure}

Taking distance estimates into account, the detection rate of X-ray emission from WD symbiotics with {\it Swift}/XRT indicates that faint X-ray emission is a common, but not universal, property of symbiotic stars. Approximately 50\% of the WD symbiotics have X-ray luminosities $L_{X} \lesssim$ a few 10$^{31}$ ergs s$^{-1}$ (see Table \ref{tab:xrays} and references in Table \ref{tab:class}). Symbiotic stars constitute an important and growing population of X-ray sources (42 objects, see Table \ref{tab:class} for details), with 7 sources showing supersoft emission, 13 sources with soft thermal X-ray emission, 7 sources with soft and hard thermal X-ray emission, 6 sources with hard thermal X-ray emission, and 9 sources with hard non-thermal X-ray emission from accretion onto a neutron star. Since 6 of our 10 {\it Swift} X-ray detections came from the top 13 of the 41 on our fill-in target list (i.e., $\sim$2/3 of our detections are from the top third of our list) that was sorted by a rough distance estimate, some of the non-detected objects (which have count rates of less than 0.0016 counts s$^{-1}$) could have similar X-ray emission to the detections, but could simply be farther away. For example, the $\beta/\delta$ system CH~Cyg would be categorized as $\delta$-type if it were 10 times farther away and were observed during a high-flux state (the distance of CH~Cyg is 245 pc and its flux varies between high and low states by factors of $\lesssim$10 below 
2 keV and approximately 30 in the 3-10 keV band; \citeads[see Table 1 in][] {2007PASJ...59S.177M}). During a low-flux state it would not be detected in a survey like the one presented here. The distance, however, cannot be the only factor in whether or not the XRT detected a source because we did not detect \object{CI~Cyg} and \object{RW~Hya}, for example, with distances of 1.5 and 0.68 kpc, respectively \citepads{1991A&A...248..458M}. 

\subsection{$\beta$-type emission}

Of the two scenarios that have been put forward as the origin of soft X-ray emission in the $\beta$- and $\beta/\delta$-type systems, we prefer the scenario that invokes shock-heated plasma due to colliding winds. Two models have been proposed to explain the $\beta$-type emission from WD symbiotics: 1) colliding winds from the outbursting WD and red giant \citepads{1997A&A...319..201M} and 2) scattering of hard X-ray photons from near the surface of the WD into our line of sight \citepads{2006MNRAS.372.1602W}. The scattering model proposed by \citetads{2006MNRAS.372.1602W} required that the binary be seen almost edge-on. Since it is unlikely that all four of the new two-component WD symbiotics are edge-on, the {\it Swift} data support the colliding winds model over the scattering model for the $\beta$ component in the X-ray spectra of WD symbiotics. The colliding winds model, as introduced by \citetads{1997A&A...319..201M}, requires the systems to be in an outburst state, so a wind from the WD can be driven. As none of our newly discovered sources seem to have experienced, to our knowledge, a recent outburst, a colliding winds model would also seem unlikely. However, the $\beta$-type WD symbiotic \object{EG~And} was detected by ROSAT, although there was no evidence of any outburst. Futhermore, \citetads{1993A&A...274L..21V} found that the observed UV line profiles can be explained by a wind from the WD in quiescence. We acknowledge that the evidence of winds from a quiescent 
WD comes from only one object (EG~And) so far; however, some (or maybe all) the sources with $\beta$-type X-ray emission in our sample have a colliding-winds region with the WD being in quiescence, constituing an increase in the previous single-object sample. 

Another possibility, not yet fully explored in symbiotic stars, is the presence of a wind from the accretion disk itself. High accretion rate, non-magnetic CVs (dwarf novae in outburst and nova-like systems) usually have an accretion disk wind \citepads{1997ASPC..121..465D}. It is likely to be line-driven, by analogy with O stars \citepads{1975ApJ...195..157C}, where those stars with luminosity above $\approx$0.1\% Eddington (where electron scattering force balances gravity) show outflows, because line scattering is $\approx$1,000 times more effective (a factor known as force multiplier) than electron scattering. \citetads{1998MNRAS.295..595P} show that CV and protostar disks with luminosity effectively above Eddington (i.e., considering the force multiplier) will drive winds, and we have no reasons to think this would be different for symbiotic stars, although this will benefit greatly from theoretical models of accretion disk in symbiotic stars.

The luminosity from this colliding wind region for the new objects and \object{CH~Cyg}, a well-known $\beta$-type system, are all commensurate (see \citeads{2007PASJ...59S.177M}) with $L_{X}[0.3-2.4$ keV]$\sim$10$^{30-31}$ ergs s$^{-1}$. The temperatures obtained from spectral models of the soft component (a few tenths of a keV) suggest plasma heated by shocks at speeds of a few hundred km s$^{-1}$ (assuming strong conditions we have $T_{shock}=3\mu m_{p} v_{shock}^{2}/16k$, where $T_{shock}$ and $v_{shock}$ are the shock temperature and speed, respectively, $m_{p}$ the proton mass, and $k$ the Boltzmann constant) , which are roughly consistent with the speeds of outflows from WD symbiotics \citepads{2007ApJ...660..651N,2004ApJ...613L..61G}.

\subsection{$\delta$-type emission}

We suggest that the hard X-ray emission in $\delta$ and $\beta/\delta$ systems is from accretion rather than quasi--steady nuclear burning or colliding winds. The high level of absorption of the hard emission shows that these high-energy photons are emitted from well within the symbiotic wind nebula. The lack of any coherent modulation of the hard X-ray emission supports our idea that the hard emission is not due to magnetic accretion onto a rotating WD (albeit our data are only sensitive to pulsed fractions of more than $\approx$ 44\%; see Sect. \ref{sec:results}). The hard X-ray component of the spectrum is well-fit by thermal models with temperatures of a few keV, which are unlikely to be produced in the colliding region of low-velocity winds. The presence of variability on time scales of minutes to hours at UV wavelengths (see Fig. \ref{fig:uvot}) supports the accretion scenario over quasi-steady nuclear burning, which varies on the much longer nuclear timescale \citepads{2003ASPC..303..202S}. Figure \ref{fig:ssexp_hr} shows that sources with the hardest X-ray spectra (see Fig. \ref{fig:spec}) are also more UV variable (Fig. \ref{fig:uvot} and group at the upper-right corner in Fig. \ref{fig:ssexp_hr}), while sources with low-amplitude UV flickering tend to have relatively little emission above 2 keV (lower-left corner group in Fig. \ref{fig:ssexp_hr}). The UVOT light curve from our unintended observation of the supersoft source StH$\alpha$~32 supports 
the proposed scenario in which sources powered by nuclear shell burning do not show large amplitude flickering (see Fig. \ref{fig:uvot} and Table \ref{time}). Moreover, unlike the WD symbiotics that produce $\delta$-type X-ray emission, the X-ray faint sources SY~Mus, CI~Cyg, and RW~Hya that were observed but not detected in our fill-in program (see Table \ref{log}) all have luminous WDs \citepads[a few hundreds to thousands L$_{\odot}$;][]{1991A&A...248..458M}. Because the amount of energy released by nuclear burning material exceeds the energy released by accretion, these sources are most likely powered by nuclear-burning material on the surface of their white dwarfs, in contrast to the sources that we detect in hard X-rays, which we believe to be mostly accretion powered. If the UV flux from any WD with quasi-steady shell burning is strong enough to Compton cool the plasma in the boundary layer, that would explain the lack of $\delta$-type emission from such WD symbiotics.

The low X-ray fluxes (especially when compared to the UV fluxes) suggest that the boundary layers are predominantly optically thick in most cases, in contrast to non-magnetic CVs where the low-accretion rate systems have the highest ratio of X-ray-to-visual flux \citepads{1985ApJ...292..535P}. The UV-to-X-ray flux ratio is less than 1 from the almost entirely optically thin boundary layer in the $\delta$-type WD symbiotic \object{T~CrB}. The unabsorbed \citepads[using E(B-V)=0.15;][]{1992ApJ...393..289S} flux in the UVOT/UVM2 filter is $F_{UVM2}$=2.3$\times$10$^{-12}$ ergs cm$^{-2}$ s$^{-1}$, while the unabsorbed flux in the 0.3-10 keV from {\it Swift}/XRT data is F$_{X}$= 3.8$\times$10$^{-11}$ ergs cm$^{-2}$ s$^{-1}$ \citepads[these data are presented in][]{2009ApJ...701.1992K}. Interestingly, the UV-to-X-ray flux ratio for \object{Hen~3-461} is $\gtrsim$0.4, suggesting that most of its X-ray flux could also originate in a mostly optically thin boundary layer. Moreover, the spectral fit of a cooling flow model yielded a mass accretion rate $\dot{M}\lesssim 4\times10^{-9}$ M$_{\odot}$/yr ($d$/1 kpc)$^{2}$ (see Sect. \ref{sec:hen}), which is within the regime of optically thin boundary layer emission around a 1 M$_{\odot}$ white dwarf as computed by \citetads{1993Natur.362..820N}. The luminosity of the $\delta$ spectral components from the objects listed in Table \ref{tab:xrays} (modulo uncertainties in the distances) ranges from 10$^{31}$ ergs s$^{-1}$ (ZZ~CMi) to 10$^{32}$ 
ergs s$^{-1}$ (V347~Nor). For comparison, the luminosity from the accretion disk boundary layer in T~CrB is on the order of 7$\times$10$^{33}$ ergs s$^{-1}$ ($d$/1 kpc)$^2$ \citepads{2008ASPC..401..342L}, suggesting that in T~CrB, the fraction of the boundary layer that is optically thin is greater than in the new systems.

Given that in symbiotics the reddening is usually derived from optical spectra \citepads[e.g.,][]{2005A&A...435.1087L} and as the newly discovered WD symbiotics presented here are poorly known at optical wavelengths, reddening values are not available in the literature for most of them. We corrected the UV fluxes (see Table \ref{tab:xrays}) for reddening of \object{V347 Nor} \citepads[E(B-V)=0.92;][]{2007apn4.confE..79S}, \object{StH$\alpha$ 32} \citepads[E(B-V)$\lesssim$0.25;]{1993A&A...268..159S} and \object{BI Cru} \citepads[E(B-V)=1.24;][]{1995A&AS..111..471P}. The UV-to-X-ray flux ratios (or its lower limit in the cases where reddening values are not available) support our conclusion about the optical depth of the boundary layer.

Because most of the boundary layers appear to be optically thick, and also because inverse Compton scattering could be cooling the plasma in the boundary layers, we cannot place tight constraints on the masses of the WDs in the symbiotics with newly detected X-ray emission. The hardness of the spectrum depends on the optical depth and temperature of the X-ray emitting plasma \citepads{1982ApJS...48..239K}. The measured temperature can be smaller than the actual shock temperature if a significant portion of the boundary layer is optically thick. 

Compton cooling of the post-shock region is important in some situations, and these situations have long been explored for magnetic CVs, in which the accretion proceeds vertically. The material in the post-shock region is cooling from the shock temperature and at temperatures of around 10$^{8}$ K (a few keV), the only important opacity source is electron scattering. For example, for accretion rates of a few 10$^{16}$ gr s$^{-1}$, a white dwarf with a mass of 0.5 M$_{\odot}$ and an accretion fractional area $f \sim$ 0.001, the optical depth for electron scattering $\tau$ is about 0.3-0.4 \citepads[see, e.g.,][]{1997ApJ...474..774F}, implying that approximately 30\% of photons will Compton scatter. However, if there is a source of seed photons that are individually less energetic but more numerous than the electrons, then each electron will experience multiple Compton scattering events. \citetads{1983ApJ...268..291I} found that in order for Compton cooling to be important we need a source of seed photons (for example the WD surface itself), high local accretion rates (accretion rate per unit area), a geometry where photons cannot easily escape without interacting with electrons, and/or a high mass WD. If these conditions are met, then the post-shock plasma will cool down through Compton scattering until the density is high enough and the temperature is low enough that bremsstrahlung cooling will start to dominate. Then the temperature derived from spectral fit of a cooling 
flow model will be lower than the shock temperature, reflecting only the portion of the shock that is being cooled by bremsstrahlung emission. The same physical mechanism should also apply to the boundary layer of non-magnetic CVs and symbiotics, although a quantitative treatment is more difficult given our still limited understanding of the boundary layer.  \citetads{2011ApJ...737....7N}  found in their analysis of the quiescent X-ray emission from \object{RS~Oph}, whose white dwarf is known to be massive and accreting at a high rate, that the shock temperature derived from the X-ray fits is 10\% of what is expected from such a massive WD.  By contrast, in \object{T~CrB}, an otherwise similar system, the fit of the X-ray spectrum yields a shock temperature compatible with the WD mass \citepads{2008ASPC..401..342L}. The temperatures derived from the X-ray fits, and the UV fluxes of the new objects presented here, suggest that they could still harbor massive white dwarfs powered by accretion rather than nuclear shell burning.

The new {\it Swift}/XRT detections of WD symbiotics do, however, allow us to place rough constraints on the rate of accretion onto the WDs in these systems. Accretion theory predicts that above a certain accretion rate, the nuclear burning occurs continuously \citepads{1982ApJ...259..244I,1982ApJ...257..767F,2007ApJ...663.1269N}. If our conclusion is valid, namely that the sources with a $\delta$ component are powered by accretion rather than by nuclear shell burning, then the accretion rate in $\delta$-type systems must be below this limit (of a few 10$^{-8}$ M$_{\odot}$ yr$^{-1}$ to a few 10$^{-7}$ M$_{\odot}$ yr$^{-1}$, depending on the WD mass). If the accretion proceeds through a disk and the boundary layer is optically thick, there is also a theoretical lower limit to $\dot{M}$ \citepads[for a particular WD mass;][]{1995ApJ...442..337P}. For $M_{WD}$=1.0, 0.8, and 0.6 M$_{\odot}$, and for the X-ray emission to be from an optically thick boundary layer, these two theoretical considerations require that $\dot{M}$ is $\gtrsim$10$^{-7}$, a few $\times$10$^{-8}$, and $\sim$10$^{-8}$ M$_{\odot}$ yr$^{-1}$, respectively. The lower limit, however, suffers from theoretical uncertainties as it depends on the adopted viscosity parameter $\alpha$. A change of about 30\% in $\alpha$ implies a change of approximately a factor of 3 in the accretion rate at which the transition from optically thin to thick boundary layer occurs. Moreover, from observations of dwarf novae, \citetads{
2011PASP..123.1054F} found that for a certain $\alpha$ the optically thin to thick transition in the boundary layer does not occur at the accretion rates predicted by  \citetads{1995ApJ...442..337P}. Regardless of these uncertainties, our data suggest that mass transfer rates on the order of $\sim$10$^{-8}$ M$_{\odot}$ yr$^{-1}$ are rather common in symbiotics and consistent with expectations from the Bondi-Hoyle accretion rate of $\dot{M}_{BH} \sim 10^{-8}$ M$_{\odot}$ yr$^{-1}$ ($M$/0.6 M$_{\odot}$)$^{2}$  (7 km s$^{-1}$/$v_{\infty}$)$^{3}$, where  $v_{\infty}$ is the relative velocity of the red giant wind at the white dwarf.

\subsection{$\beta/\delta$-type X-ray spectra}

$\beta/\delta$-type X-ray spectra are present in approximately 20\% of the WD symbiotics, and could be associated with the production of bi-polar outflows. The new $\beta/\delta$ systems that we have discovered with {\it Swift}, \object{NQ~Gem}, \object{ZZ~CMi}, \object{V347 Nor}, and \object{UV Aur}, have spectra that resemble the well-known X-ray spectra from the WD symbiotics CH~Cyg and R~Aqr, suggesting that although this X-ray spectral type was previously thought to be unusual, it is actually common. If these four new objects had been observed with ROSAT, they would have been classified as $\beta$-type in the scheme of \citetads{1997A&A...319..201M}. However, these objects also display a hard X-ray component characteristic of $\delta$-type systems (see above). Therefore, we revise the \citetads{1997A&A...319..201M} classification scheme and categorize two-component X-ray spectra as $\beta/\delta$ type. 

An interesting similarity between the recently discovered $\beta/\delta$-type WD symbiotic \object{V347 Nor} and the previously known $\beta/\delta$-type \object{CH~Cyg} and \object{R~Aqr} is that all have extended bi-polar outflows \citepads[e.g.,][]{1999A&A...348..978C}. The luminosity of the $\beta$ component in our newly discovered $\beta/\delta$-type WD symbiotic, however, is higher than the luminosity (in the 0.3-1 keV energy range) of the jet components in \object{CH~Cyg} and \object{R~Aqr}. The $\beta$ components of \object{V347~Nor} has luminosity of approximately 5$\times$10$^{30}$ (d/ 1.5 kpc)$^{2}$. In turn, the jet component in CH~Cyg has a luminosity of 5$\times$10$^{28}$ ergs s$^{-1}$ (d/245 pc)$^{2}$ \citepads{2007ApJ...661.1048K}; the NE jet in R~Aqr has a luminosity of 7$\times$10$^{29}$ ergs s$^{-1}$; and the SE jet has a luminosity of 2$\times$10$^{29}$ ergs s$^{-1}$ \citepads{2007ApJ...660..651N}. Thus, the jet emission in $\beta/\delta$-type WD symbiotics is not contributing significantly to the flux of the $\beta$ component. Moreover, both CH~Cyg and R~Aqr have spatially unresolved $\beta$-type emission that is much stronger than the jet emission. Then, the $\beta$-type emission seems to be either from the inner, spatially-unresolved portions of the jet or from some other source of emission that preferentially appears when jets are present. Although not yet observed with sensitive hard X-ray detectors, our findings suggest that \object{V1016~Cyg} and \
object{HM~Sge}, both symbiotic binaries with outflows detected in optical, could also be $\beta/\delta$-type systems.

\subsection{Conclusions}

\begin{enumerate}
\item X-ray emission is a common feature of WD symbiotics. That X-rays have been preferentially detected from nearby sources, 25\% of known WD symbiotics, suggest such emission is prevalent.
\item The X-ray spectra of WD symbiotics show three distinct spectral components: $\alpha$, which is associated with quasi-steady shell burning; $\beta$, which is most likely from colliding winds; and $\delta$, which we propose is from the innermost accretion region.
$\beta$- and $\delta$-type X-ray emission are often, but not always, found together.
\item The UV-to-X-ray flux ratio of the $\delta$-type targets reveals that the innermost accretion region, which is probably a boundary layer in most cases, is often optically thick, as expected for 0.6 M$_{\odot}$ WDs accreting at the Bondi-Hoyle rate of $\approx$10$^{-8}$ M$_{\odot}$/yr.
\item Although most WD symbiotics do not produce detectable optical flickering on time scales of minutes, rapid UV flickering, presumably associated with accretion, is pervasive.
\end{enumerate}

\begin{acknowledgements}

We thank the anonymous referee for comments and suggestions which improved the final quality of this article. We acknowledge the {\it Swift} team for planning these observations. G. J. M. Luna is a member of the CIC-CONICET (Argentina) and acknowledges support from grants PIP-Conicet/2011 \#D4598 and FONCyT/PICT/2011 \#269. J. L. Sokoloski acknowledges support from grants SAO GO1-12041A, NASA NNX10AK31G, and NASA NNX11AD77G. We thank N. Masetti and P. Evans for useful comments that helped to improve the manuscript.
\end{acknowledgements}
\bibliographystyle{aa}    
\bibliography{listaref}

\begin{thebibliography}{109}
\expandafter\ifx\csname natexlab\endcsname\relax\def\natexlab#1{#1}\fi

\bibitem[{{Ake}(1979)}]{1979ApJ...234..538A}
{Ake}, T.~B. 1979, \apj, 234, 538

\bibitem[{{Alexander} {et~al.}(2011){Alexander}, {Wynn}, {King}, \&
  {Pringle}}]{2011MNRAS.418.2576A}
{Alexander}, R.~D., {Wynn}, G.~A., {King}, A.~R., \& {Pringle}, J.~E. 2011,
  \mnras, 418, 2576

\bibitem[{{Allen}(1984)}]{1984PASAu...5..369A}
{Allen}, D.~A. 1984, Proceedings of the Astronomical Society of Australia, 5,
  369

\bibitem[{{Anders} \& {Grevesse}(1989)}]{1989GeCoA..53..197A}
{Anders}, E. \& {Grevesse}, N. 1989, \gca, 53, 197

\bibitem[{{Arnaud} {et~al.}(2011){Arnaud}, {Smith}, \&
  {Siemiginowska}}]{2011hxa..book.....A}
{Arnaud}, K., {Smith}, R., \& {Siemiginowska}, A. 2011, {Handbook of X-ray
  Astronomy}, ed. R.~{Ellis}, J.~{Huchra}, S.~{Kahn}, G.~{Rieke}, \& P.~B.
  {Stetson}

\bibitem[{{Arnaud}(1996)}]{1996ASPC..101...17A}
{Arnaud}, K.~A. 1996, in Astronomical Society of the Pacific Conference Series,
  Vol. 101, Astronomical Data Analysis Software and Systems V, ed. G.~H.
  {Jacoby} \& J.~{Barnes}, 17

\bibitem[{{Baumgartner} {et~al.}(2012){Baumgartner}, {Tueller}, {Markwardt},
  {Skinner}, {Barthelmy}, {Mushotzky}, {Evans}, \&
  {Gehrels}}]{2012arXiv1212.3336B}
{Baumgartner}, W.~H., {Tueller}, J., {Markwardt}, C.~B., {et~al.} 2012, ArXiv
  e-prints

\bibitem[{{Belczy{\'n}ski} {et~al.}(2000){Belczy{\'n}ski}, {Miko{\l}ajewska},
  {Munari}, {Ivison}, \& {Friedjung}}]{2000A&AS..146..407B}
{Belczy{\'n}ski}, K., {Miko{\l}ajewska}, J., {Munari}, U., {Ivison}, R.~J., \&
  {Friedjung}, M. 2000, \aaps, 146, 407

\bibitem[{{Bickert} {et~al.}(1996){Bickert}, {Greiner}, \&
  {Stencel}}]{1996LNP...472..225B}
{Bickert}, K.~F., {Greiner}, J., \& {Stencel}, R.~E. 1996, in Lecture Notes in
  Physics, Berlin Springer Verlag, Vol. 472, Supersoft X-Ray Sources, ed.
  J.~{Greiner}, 225

\bibitem[{{Bondi} \& {Hoyle}(1944)}]{1944MNRAS.104..273B}
{Bondi}, H. \& {Hoyle}, F. 1944, \mnras, 104, 273

\bibitem[{{Buccheri} {et~al.}(1983){Buccheri}, {Bennett}, {Bignami}, {Bloemen},
  {Boriakoff}, {Caraveo}, {Hermsen}, {Kanbach}, {Manchester}, {Masnou},
  {Mayer-Hasselwander}, {Ozel}, {Paul}, {Sacco}, {Scarsi}, \&
  {Strong}}]{1983A&A...128..245B}
{Buccheri}, R., {Bennett}, K., {Bignami}, G.~F., {et~al.} 1983, \aap, 128, 245

\bibitem[{{Byckling} {et~al.}(2010){Byckling}, {Mukai}, {Thorstensen}, \&
  {Osborne}}]{2010MNRAS.408.2298B}
{Byckling}, K., {Mukai}, K., {Thorstensen}, J.~R., \& {Osborne}, J.~P. 2010,
  \mnras, 408, 2298

\bibitem[{{Carquillat} \& {Prieur}(2008)}]{2008AN....329...44C}
{Carquillat}, J.-M. \& {Prieur}, J.-L. 2008, Astronomische Nachrichten, 329, 44

\bibitem[{{Cash}(1979)}]{1979ApJ...228..939C}
{Cash}, W. 1979, \apj, 228, 939

\bibitem[{{Castor} {et~al.}(1975){Castor}, {Abbott}, \&
  {Klein}}]{1975ApJ...195..157C}
{Castor}, J.~I., {Abbott}, D.~C., \& {Klein}, R.~I. 1975, \apj, 195, 157

\bibitem[{{Chaty} {et~al.}(2008){Chaty}, {Rahoui}, {Foellmi}, {Tomsick},
  {Rodriguez}, \& {Walter}}]{2008A&A...484..783C}
{Chaty}, S., {Rahoui}, F., {Foellmi}, C., {et~al.} 2008, \aap, 484, 783

\bibitem[{{Chernyakova} {et~al.}(2005){Chernyakova}, {Courvoisier},
  {Rodriguez}, \& {Lutovinov}}]{2005ATel..519....1C}
{Chernyakova}, M., {Courvoisier}, T.~J.-L., {Rodriguez}, J., \& {Lutovinov}, A.
  2005, The Astronomer's Telegram, 519, 1

\bibitem[{{Chiotellis} {et~al.}(2012){Chiotellis}, {Schure}, \&
  {Vink}}]{2012A&A...537A.139C}
{Chiotellis}, A., {Schure}, K.~M., \& {Vink}, J. 2012, \aap, 537, A139

\bibitem[{{Cieslinski} {et~al.}(1994){Cieslinski}, {Elizalde}, \&
  {Steiner}}]{1994A&AS..106..243C}
{Cieslinski}, D., {Elizalde}, F., \& {Steiner}, J.~E. 1994, \aaps, 106, 243

\bibitem[{{Contini} {et~al.}(2009){Contini}, {Angeloni}, \&
  {Rafanelli}}]{2009MNRAS.396..807C}
{Contini}, M., {Angeloni}, R., \& {Rafanelli}, P. 2009, \mnras, 396, 807

\bibitem[{{Corbet} {et~al.}(2008){Corbet}, {Sokoloski}, {Mukai}, {Markwardt},
  \& {Tueller}}]{2008ApJ...675.1424C}
{Corbet}, R.~H.~D., {Sokoloski}, J.~L., {Mukai}, K., {Markwardt}, C.~B., \&
  {Tueller}, J. 2008, \apj, 675, 1424

\bibitem[{{Corradi} {et~al.}(1999){Corradi}, {Ferrer}, {Schwarz}, {Brandi}, \&
  {Garc{\'{\i}}a}}]{1999A&A...348..978C}
{Corradi}, R.~L.~M., {Ferrer}, O.~E., {Schwarz}, H.~E., {Brandi}, E., \&
  {Garc{\'{\i}}a}, L. 1999, \aap, 348, 978

\bibitem[{{Corradi} \& {Schwarz}(1993)}]{1993A&A...268..714C}
{Corradi}, R.~L.~M. \& {Schwarz}, H.~E. 1993, \aap, 268, 714

\bibitem[{{Di Stefano}(2010)}]{2010ApJ...719..474D}
{Di Stefano}, R. 2010, \apj, 719, 474

\bibitem[{{Dilday} {et~al.}(2012){Dilday}, {Howell}, {Cenko}, {Silverman},
  {Nugent}, {Sullivan}, {Ben-Ami}, {Bildsten}, {Bolte}, {Endl}, {Filippenko},
  {Gnat}, {Horesh}, {Hsiao}, {Kasliwal}, {Kirkman}, {Maguire}, {Marcy},
  {Moore}, {Pan}, {Parrent}, {Podsiadlowski}, {Quimby}, {Sternberg}, {Suzuki},
  {Tytler}, {Xu}, {Bloom}, {Gal-Yam}, {Hook}, {Kulkarni}, {Law}, {Ofek},
  {Polishook}, \& {Poznanski}}]{2012Sci...337..942D}
{Dilday}, B., {Howell}, D.~A., {Cenko}, S.~B., {et~al.} 2012, Science, 337, 942

\bibitem[{{Drew}(1997)}]{1997ASPC..121..465D}
{Drew}, J.~E. 1997, in Astronomical Society of the Pacific Conference Series,
  Vol. 121, IAU Colloq. 163: Accretion Phenomena and Related Outflows, ed.
  D.~T. {Wickramasinghe}, G.~V. {Bicknell}, \& L.~{Ferrario}, 465

\bibitem[{{Eze} {et~al.}(2010){Eze}, {Luna}, \& {Smith}}]{2010ApJ...709..816E}
{Eze}, R.~N.~C., {Luna}, G.~J.~M., \& {Smith}, R.~K. 2010, \apj, 709, 816

\bibitem[{{Fertig} {et~al.}(2011){Fertig}, {Mukai}, {Nelson}, \&
  {Cannizzo}}]{2011PASP..123.1054F}
{Fertig}, D., {Mukai}, K., {Nelson}, T., \& {Cannizzo}, J.~K. 2011, \pasp, 123,
  1054

\bibitem[{{Fujimoto}(1982)}]{1982ApJ...257..767F}
{Fujimoto}, M.~Y. 1982, \apj, 257, 767

\bibitem[{{Fujimoto} \& {Ishida}(1997)}]{1997ApJ...474..774F}
{Fujimoto}, R. \& {Ishida}, M. 1997, \apj, 474, 774

\bibitem[{{Galloway} \& {Sokoloski}(2004)}]{2004ApJ...613L..61G}
{Galloway}, D.~K. \& {Sokoloski}, J.~L. 2004, \apjl, 613, L61

\bibitem[{{Gon{\c c}alves} {et~al.}(2008){Gon{\c c}alves}, {Magrini}, {Munari},
  {Corradi}, \& {Costa}}]{2008MNRAS.391L..84G}
{Gon{\c c}alves}, D.~R., {Magrini}, L., {Munari}, U., {Corradi}, R.~L.~M., \&
  {Costa}, R.~D.~D. 2008, \mnras, 391, L84

\bibitem[{{Greene} \& {Wing}(1971)}]{1971ApJ...163..309G}
{Greene}, A.~E. \& {Wing}, R.~F. 1971, \apj, 163, 309

\bibitem[{{Haakonsen} \& {Rutledge}(2009)}]{2009ApJS..184..138H}
{Haakonsen}, C.~B. \& {Rutledge}, R.~E. 2009, \apjs, 184, 138

\bibitem[{{Herbig}(2009)}]{2009AJ....138.1502H}
{Herbig}, G.~H. 2009, \aj, 138, 1502

\bibitem[{{Iben}(1982)}]{1982ApJ...259..244I}
{Iben}, Jr., I. 1982, \apj, 259, 244

\bibitem[{{Ikeda} \& {Tamura}(2004)}]{2004PASJ...56..353I}
{Ikeda}, Y. \& {Tamura}, S. 2004, \pasj, 56, 353

\bibitem[{{Imamura} \& {Durisen}(1983)}]{1983ApJ...268..291I}
{Imamura}, J.~N. \& {Durisen}, R.~H. 1983, \apj, 268, 291

\bibitem[{{Ishida} {et~al.}(2009){Ishida}, {Okada}, {Hayashi}, {Nakamura},
  {Terada}, {Mukai}, \& {Hamaguchi}}]{2009PASJ...61S..77I}
{Ishida}, M., {Okada}, S., {Hayashi}, T., {et~al.} 2009, \pasj, 61, 77

\bibitem[{{Jorissen} {et~al.}(2012){Jorissen}, {Van Eck}, {Dermine}, {Van
  Winckel}, \& {Gorlova}}]{2012BaltA..21...39J}
{Jorissen}, A., {Van Eck}, S., {Dermine}, T., {Van Winckel}, H., \& {Gorlova},
  N. 2012, Baltic Astronomy, 21, 39

\bibitem[{{Kaplan} {et~al.}(2007){Kaplan}, {Levine}, {Chakrabarty}, {Morgan},
  {Erb}, {Gaensler}, {Moon}, \& {Cameron}}]{2007ApJ...661..437K}
{Kaplan}, D.~L., {Levine}, A.~M., {Chakrabarty}, D., {et~al.} 2007, \apj, 661,
  437

\bibitem[{{Karovska} {et~al.}(2007){Karovska}, {Carilli}, {Raymond}, \&
  {Mattei}}]{2007ApJ...661.1048K}
{Karovska}, M., {Carilli}, C.~L., {Raymond}, J.~C., \& {Mattei}, J.~A. 2007,
  \apj, 661, 1048

\bibitem[{{Karovska} {et~al.}(2005){Karovska}, {Schlegel}, {Hack}, {Raymond},
  \& {Wood}}]{2005ApJ...623L.137K}
{Karovska}, M., {Schlegel}, E., {Hack}, W., {Raymond}, J.~C., \& {Wood}, B.~E.
  2005, \apjl, 623, L137

\bibitem[{{Kennea} {et~al.}(2009){Kennea}, {Mukai}, {Sokoloski}, {Luna},
  {Tueller}, {Markwardt}, \& {Burrows}}]{2009ApJ...701.1992K}
{Kennea}, J.~A., {Mukai}, K., {Sokoloski}, J.~L., {et~al.} 2009, \apj, 701,
  1992

\bibitem[{{Kenny} \& {Taylor}(2005)}]{2005ApJ...619..527K}
{Kenny}, H.~T. \& {Taylor}, A.~R. 2005, \apj, 619, 527

\bibitem[{{Kylafis} \& {Lamb}(1982)}]{1982ApJS...48..239K}
{Kylafis}, N.~D. \& {Lamb}, D.~Q. 1982, \apjs, 48, 239

\bibitem[{{Livio} \& {Warner}(1984)}]{1984Obs...104..152L}
{Livio}, M. \& {Warner}, B. 1984, The Observatory, 104, 152

\bibitem[{{Luna} \& {Costa}(2005)}]{2005A&A...435.1087L}
{Luna}, G.~J.~M. \& {Costa}, R.~D.~D. 2005, \aap, 435, 1087

\bibitem[{{Luna} \& {Sokoloski}(2007)}]{2007ApJ...671..741L}
{Luna}, G.~J.~M. \& {Sokoloski}, J.~L. 2007, \apj, 671, 741

\bibitem[{{Luna} {et~al.}(2006){Luna}, {Sokoloski}, \&
  {Costa}}]{2006Ap&SS.304..283L}
{Luna}, G.~J.~M., {Sokoloski}, J.~L., \& {Costa}, R.~D.~D. 2006, \apss, 304,
  283

\bibitem[{{Luna} {et~al.}(2008){Luna}, {Sokoloski}, \&
  {Mukai}}]{2008ASPC..401..342L}
{Luna}, G.~J.~M., {Sokoloski}, J.~L., \& {Mukai}, K. 2008, in Astronomical
  Society of the Pacific Conference Series, Vol. 401, RS Ophiuchi (2006) and
  the Recurrent Nova Phenomenon, ed. A.~{Evans}, M.~F. {Bode}, T.~J. {O'Brien},
  \& M.~J. {Darnley}, 342

\bibitem[{{Luna} {et~al.}(2012){Luna}, {Sokoloski}, {Mukai}, \&
  {Nunez}}]{2012ATel.3960....1L}
{Luna}, G.~J.~M., {Sokoloski}, J.~L., {Mukai}, K., \& {Nunez}, N. 2012, The
  Astronomer's Telegram, 3960, 1

\bibitem[{{Marcu} {et~al.}(2011){Marcu}, {F{\"u}rst}, {Pottschmidt},
  {Grinberg}, {M{\"u}ller}, {Wilms}, {Postnov}, {Corbet}, {Markwardt}, \&
  {Cadolle Bel}}]{2011ApJ...742L..11M}
{Marcu}, D.~M., {F{\"u}rst}, F., {Pottschmidt}, K., {et~al.} 2011, \apjl, 742,
  L11

\bibitem[{{Masetti} {et~al.}(2005){Masetti}, {Bassani}, {Bird}, \&
  {Bazzano}}]{2005ATel..528....1M}
{Masetti}, N., {Bassani}, L., {Bird}, A.~J., \& {Bazzano}, A. 2005, The
  Astronomer's Telegram, 528, 1

\bibitem[{{Masetti} {et~al.}(2002){Masetti}, {Dal Fiume}, {Cusumano}, {Amati},
  {Bartolini}, {Del Sordo}, {Frontera}, {Guarnieri}, {Orlandini}, {Palazzi},
  {Parmar}, {Piccioni}, \& {Santangelo}}]{2002A&A...382..104M}
{Masetti}, N., {Dal Fiume}, D., {Cusumano}, G., {et~al.} 2002, \aap, 382, 104

\bibitem[{{Masetti} {et~al.}(2007{\natexlab{a}}){Masetti}, {Landi},
  {Pretorius}, {Sguera}, {Bird}, {Perri}, {Charles}, {Kennea}, {Malizia}, \&
  {Ubertini}}]{2007A&A...470..331M}
{Masetti}, N., {Landi}, R., {Pretorius}, M.~L., {et~al.} 2007{\natexlab{a}},
  \aap, 470, 331

\bibitem[{{Masetti} {et~al.}(2011){Masetti}, {Munari}, {Henden}, {Page},
  {Osborne}, \& {Starrfield}}]{2011A&A...534A..89M}
{Masetti}, N., {Munari}, U., {Henden}, A.~A., {et~al.} 2011, \aap, 534, A89

\bibitem[{{Masetti} {et~al.}(2006){Masetti}, {Orlandini}, {Palazzi}, {Amati},
  \& {Frontera}}]{2006A&A...453..295M}
{Masetti}, N., {Orlandini}, M., {Palazzi}, E., {Amati}, L., \& {Frontera}, F.
  2006, \aap, 453, 295

\bibitem[{{Masetti} {et~al.}(2012){Masetti}, {Parisi},
  {Jim{\'e}nez-Bail{\'o}n}, {Palazzi}, {Chavushyan}, {Bassani}, {Bazzano},
  {Bird}, {Dean}, {Galaz}, {Landi}, {Malizia}, {Minniti}, {Morelli},
  {Schiavone}, {Stephen}, \& {Ubertini}}]{2012A&A...538A.123M}
{Masetti}, N., {Parisi}, P., {Jim{\'e}nez-Bail{\'o}n}, E., {et~al.} 2012, \aap,
  538, A123

\bibitem[{{Masetti} {et~al.}(2007{\natexlab{b}}){Masetti}, {Rigon}, {Maiorano},
  {Cusumano}, {Palazzi}, {Orlandini}, {Amati}, \&
  {Frontera}}]{2007A&A...464..277M}
{Masetti}, N., {Rigon}, E., {Maiorano}, E., {et~al.} 2007{\natexlab{b}}, \aap,
  464, 277

\bibitem[{{McCollum} {et~al.}(2008){McCollum}, {Bruhweiler}, {Wahlgren},
  {Eriksson}, \& {Verner}}]{2008ApJ...682.1087M}
{McCollum}, B., {Bruhweiler}, F.~C., {Wahlgren}, G.~M., {Eriksson}, M., \&
  {Verner}, E. 2008, \apj, 682, 1087

\bibitem[{{Miko{\l}ajewska}(2007)}]{2007BaltA..16....1M}
{Miko{\l}ajewska}, J. 2007, Baltic Astronomy, 16, 1

\bibitem[{{Montez} {et~al.}(2006){Montez}, {Kastner}, \&
  {Sahai}}]{2006AAS...209.9206M}
{Montez}, Jr., R., {Kastner}, J.~H., \& {Sahai}, R. 2006, in Bulletin of the
  American Astronomical Society, Vol.~38, American Astronomical Society Meeting
  Abstracts, 1029

\bibitem[{{Muerset} {et~al.}(1991){Muerset}, {Nussbaumer}, {Schmid}, \&
  {Vogel}}]{1991A&A...248..458M}
{Muerset}, U., {Nussbaumer}, H., {Schmid}, H.~M., \& {Vogel}, M. 1991, \aap,
  248, 458

\bibitem[{{Muerset} {et~al.}(1997){Muerset}, {Wolff}, \&
  {Jordan}}]{1997A&A...319..201M}
{Muerset}, U., {Wolff}, B., \& {Jordan}, S. 1997, \aap, 319, 201

\bibitem[{{Mukai} {et~al.}(2007){Mukai}, {Ishida}, {Kilbourne}, {Mori},
  {Terada}, {Chan}, \& {Soong}}]{2007PASJ...59S.177M}
{Mukai}, K., {Ishida}, M., {Kilbourne}, C., {et~al.} 2007, \pasj, 59, 177

\bibitem[{{Munari} \& {Patat}(1993)}]{1993A&A...277..195M}
{Munari}, U. \& {Patat}, F. 1993, \aap, 277, 195

\bibitem[{{Munari} \& {Renzini}(1992)}]{1992ApJ...397L..87M}
{Munari}, U. \& {Renzini}, A. 1992, \apjl, 397, L87

\bibitem[{{Narayan} \& {Popham}(1993)}]{1993Natur.362..820N}
{Narayan}, R. \& {Popham}, R. 1993, \nat, 362, 820

\bibitem[{{Nelson} {et~al.}(2011){Nelson}, {Mukai}, {Orio}, {Luna}, \&
  {Sokoloski}}]{2011ApJ...737....7N}
{Nelson}, T., {Mukai}, K., {Orio}, M., {Luna}, G.~J.~M., \& {Sokoloski}, J.~L.
  2011, \apj, 737, 7

\bibitem[{{Nespoli} {et~al.}(2010){Nespoli}, {Fabregat}, \&
  {Mennickent}}]{2010A&A...516A..94N}
{Nespoli}, E., {Fabregat}, J., \& {Mennickent}, R.~E. 2010, \aap, 516, A94

\bibitem[{{Nichols} {et~al.}(2007){Nichols}, {DePasquale}, {Kellogg},
  {Anderson}, {Sokoloski}, \& {Pedelty}}]{2007ApJ...660..651N}
{Nichols}, J.~S., {DePasquale}, J., {Kellogg}, E., {et~al.} 2007, \apj, 660,
  651

\bibitem[{{Nomoto} {et~al.}(2007){Nomoto}, {Saio}, {Kato}, \&
  {Hachisu}}]{2007ApJ...663.1269N}
{Nomoto}, K., {Saio}, H., {Kato}, M., \& {Hachisu}, I. 2007, \apj, 663, 1269

\bibitem[{{Orio} {et~al.}(2012){Orio}, {Behar}, {Gallagher}, {Bianchini},
  {Chiosi}, {Luna}, {Nelson}, {Rauch}, {Schaefer}, \&
  {Tofflemire}}]{2012MNRAS.tmp..361O}
{Orio}, M., {Behar}, E., {Gallagher}, J., {et~al.} 2012, \mnras, 361

\bibitem[{{Orio} {et~al.}(2007){Orio}, {Zezas}, {Munari}, {Siviero}, \&
  {Tepedelenlioglu}}]{2007ApJ...661.1105O}
{Orio}, M., {Zezas}, A., {Munari}, U., {Siviero}, A., \& {Tepedelenlioglu}, E.
  2007, \apj, 661, 1105

\bibitem[{{Patat} {et~al.}(2007){Patat}, {Chandra}, {Chevalier}, {Justham},
  {Podsiadlowski}, {Wolf}, {Gal-Yam}, {Pasquini}, {Crawford}, {Mazzali},
  {Pauldrach}, {Nomoto}, {Benetti}, {Cappellaro}, {Elias-Rosa}, {Hillebrandt},
  {Leonard}, {Pastorello}, {Renzini}, {Sabbadin}, {Simon}, \&
  {Turatto}}]{2007Sci...317..924P}
{Patat}, F., {Chandra}, P., {Chevalier}, R., {et~al.} 2007, Science, 317, 924

\bibitem[{{Patterson} \& {Raymond}(1985)}]{1985ApJ...292..535P}
{Patterson}, J. \& {Raymond}, J.~C. 1985, \apj, 292, 535

\bibitem[{{Pereira}(1995)}]{1995A&AS..111..471P}
{Pereira}, C.~B. 1995, \aaps, 111, 471

\bibitem[{{Pereira} {et~al.}(1998){Pereira}, {Landaberry}, \& {da Concei{\c
  c}{\~a}o}}]{1998AJ....116.1971P}
{Pereira}, C.~B., {Landaberry}, S.~J.~C., \& {da Concei{\c c}{\~a}o}, F. 1998,
  \aj, 116, 1971

\bibitem[{{Podsiadlowski} \& {Mohamed}(2007)}]{2007BaltA..16...26P}
{Podsiadlowski}, P. \& {Mohamed}, S. 2007, Baltic Astronomy, 16, 26

\bibitem[{{Poole} {et~al.}(2008){Poole}, {Breeveld}, {Page}, {Landsman},
  {Holland}, {Roming}, {Kuin}, {Brown}, {Gronwall}, {Hunsberger}, {Koch},
  {Mason}, {Schady}, {vanden Berk}, {Blustin}, {Boyd}, {Broos}, {Carter},
  {Chester}, {Cucchiara}, {Hancock}, {Huckle}, {Immler}, {Ivanushkina},
  {Kennedy}, {Marshall}, {Morgan}, {Pandey}, {de Pasquale}, {Smith}, \&
  {Still}}]{2008MNRAS.383..627P}
{Poole}, T.~S., {Breeveld}, A.~A., {Page}, M.~J., {et~al.} 2008, \mnras, 383,
  627

\bibitem[{{Popham} \& {Narayan}(1995)}]{1995ApJ...442..337P}
{Popham}, R. \& {Narayan}, R. 1995, \apj, 442, 337

\bibitem[{{Proga} {et~al.}(1998){Proga}, {Stone}, \&
  {Drew}}]{1998MNRAS.295..595P}
{Proga}, D., {Stone}, J.~M., \& {Drew}, J.~E. 1998, \mnras, 295, 595

\bibitem[{{Protassov} {et~al.}(2002){Protassov}, {van Dyk}, {Connors},
  {Kashyap}, \& {Siemiginowska}}]{2002ApJ...571..545P}
{Protassov}, R., {van Dyk}, D.~A., {Connors}, A., {Kashyap}, V.~L., \&
  {Siemiginowska}, A. 2002, \apj, 571, 545

\bibitem[{{Sanford}(1949)}]{1949PASP...61..261S}
{Sanford}, R.~F. 1949, \pasp, 61, 261

\bibitem[{{Sanford}(1950)}]{1950ApJ...111..270S}
{Sanford}, R.~F. 1950, \apj, 111, 270

\bibitem[{{Santander-Garc{\'{\i}}a} {et~al.}(2009){Santander-Garc{\'{\i}}a},
  {Corradi}, \& {Mampaso}}]{2007apn4.confE..79S}
{Santander-Garc{\'{\i}}a}, M., {Corradi}, R.~L.~M., \& {Mampaso}, A. 2009, in
  Asymmetrical Planetary Nebulae IV

\bibitem[{{Santander-Garc{\'{\i}}a} {et~al.}(2007){Santander-Garc{\'{\i}}a},
  {Corradi}, {Whitelock}, {Munari}, {Mampaso}, {Marang}, {Boffi}, \&
  {Livio}}]{2007A&A...465..481S}
{Santander-Garc{\'{\i}}a}, M., {Corradi}, R.~L.~M., {Whitelock}, P.~A.,
  {et~al.} 2007, \aap, 465, 481

\bibitem[{{Schmid}(1994)}]{1994A&A...284..156S}
{Schmid}, H.~M. 1994, \aap, 284, 156

\bibitem[{{Schmid} \& {Nussbaumer}(1993)}]{1993A&A...268..159S}
{Schmid}, H.~M. \& {Nussbaumer}, H. 1993, \aap, 268, 159

\bibitem[{{Seal}(1988)}]{1988syph.book..293S}
{Seal}, P. 1988, {The Symbiotic Star UV Aurigae}, ed. J.~{Mikolajewska},
  M.~{Friedjung}, S.~J. {Kenyon}, \& R.~{Viotti}, 293

\bibitem[{{Selvelli} {et~al.}(1992){Selvelli}, {Cassatella}, \&
  {Gilmozzi}}]{1992ApJ...393..289S}
{Selvelli}, P.~L., {Cassatella}, A., \& {Gilmozzi}, R. 1992, \apj, 393, 289

\bibitem[{{Smith} {et~al.}(2008){Smith}, {Mushotzky}, {Mukai}, {Kallman},
  {Markwardt}, \& {Tueller}}]{2008PASJ...60S..43S}
{Smith}, R.~K., {Mushotzky}, R., {Mukai}, K., {et~al.} 2008, \pasj, 60, 43

\bibitem[{{Sokoloski}(2003)}]{2003ASPC..303..202S}
{Sokoloski}, J.~L. 2003, in Astronomical Society of the Pacific Conference
  Series, Vol. 303, Symbiotic Stars Probing Stellar Evolution, ed. R.~L.~M.
  {Corradi}, J.~{Mikolajewska}, \& T.~J. {Mahoney}, 202

\bibitem[{{Sokoloski} \& {Bildsten}(2010)}]{2010ApJ...723.1188S}
{Sokoloski}, J.~L. \& {Bildsten}, L. 2010, \apj, 723, 1188

\bibitem[{{Sokoloski} {et~al.}(2006{\natexlab{a}}){Sokoloski}, {Kenyon},
  {Espey}, {Keyes}, {McCandliss}, {Kong}, {Aufdenberg}, {Filippenko}, {Li},
  {Brocksopp}, {Kaiser}, {Charles}, {Rupen}, \& {Stone}}]{2006ApJ...636.1002S}
{Sokoloski}, J.~L., {Kenyon}, S.~J., {Espey}, B.~R., {et~al.}
  2006{\natexlab{a}}, \apj, 636, 1002

\bibitem[{{Sokoloski} {et~al.}(2006{\natexlab{b}}){Sokoloski}, {Luna}, {Mukai},
  \& {Kenyon}}]{2006Natur.442..276S}
{Sokoloski}, J.~L., {Luna}, G.~J.~M., {Mukai}, K., \& {Kenyon}, S.~J.
  2006{\natexlab{b}}, \nat, 442, 276

\bibitem[{{Stute} {et~al.}(2013){Stute}, {Luna}, {Pillitteri}, \&
  {Sokoloski}}]{2013A&A...554A..56S}
{Stute}, M., {Luna}, G.~J.~M., {Pillitteri}, I.~F., \& {Sokoloski}, J.~L. 2013,
  \aap, 554, A56

\bibitem[{{Stute} {et~al.}(2011){Stute}, {Luna}, \&
  {Sokoloski}}]{2011ApJ...731...12S}
{Stute}, M., {Luna}, G.~J.~M., \& {Sokoloski}, J.~L. 2011, \apj, 731, 12

\bibitem[{{Stute} \& {Sahai}(2009)}]{2009A&A...498..209S}
{Stute}, M. \& {Sahai}, R. 2009, \aap, 498, 209

\bibitem[{{Thompson} {et~al.}(2006){Thompson}, {Tomsick}, {Rothschild}, {in't
  Zand}, \& {Walter}}]{2006ApJ...649..373T}
{Thompson}, T.~W.~J., {Tomsick}, J.~A., {Rothschild}, R.~E., {in't Zand},
  J.~J.~M., \& {Walter}, R. 2006, \apj, 649, 373

\bibitem[{{Tueller} {et~al.}(2005){Tueller}, {Gehrels}, {Mushotzky},
  {Markwardt}, {Kennea}, {Burrows}, {Mukai}, \&
  {Sokoloski}}]{2005ATel..591....1T}
{Tueller}, J., {Gehrels}, N., {Mushotzky}, R.~F., {et~al.} 2005, The
  Astronomer's Telegram, 591, 1

\bibitem[{{Van Eck} \& {Jorissen}(1999)}]{1999A&A...345..127V}
{Van Eck}, S. \& {Jorissen}, A. 1999, \aap, 345, 127

\bibitem[{{Van Eck} \& {Jorissen}(2002)}]{2002A&A...396..599V}
{Van Eck}, S. \& {Jorissen}, A. 2002, \aap, 396, 599

\bibitem[{{Vogel}(1993)}]{1993A&A...274L..21V}
{Vogel}, M. 1993, \aap, 274, L21

\bibitem[{{Wang} \& {Han}(2010)}]{2010RAA....10..235W}
{Wang}, B. \& {Han}, Z.-W. 2010, Research in Astronomy and Astrophysics, 10,
  235

\bibitem[{{Warner}(1995)}]{1995CAS....28.....W}
{Warner}, B. 1995, Cambridge Astrophysics Series, 28

\bibitem[{{Wheatley} \& {Kallman}(2006)}]{2006MNRAS.372.1602W}
{Wheatley}, P.~J. \& {Kallman}, T.~R. 2006, \mnras, 372, 1602

\bibitem[{{Wynn}(2008)}]{2008ASPC..401...73W}
{Wynn}, G. 2008, in Astronomical Society of the Pacific Conference Series, Vol.
  401, RS Ophiuchi (2006) and the Recurrent Nova Phenomenon, ed. A.~{Evans},
  M.~F. {Bode}, T.~J. {O'Brien}, \& M.~J. {Darnley}, 73

\end{thebibliography}

\longtab{3}{
\begin{longtable}{lccccc}
\caption{UVOT Timing analysis results. For those objects observed in more than one visit, we list the standard deviations, $s$ and $s_{exp}$ for each visit that contains more than two exposures. The mean count rate during each visit is listed under the column $<$ {\it count rate} $>$, while $s_{frac}$ represents the fractional rms variability amplitudes that we define as $s$/$<$ {\it count rate} $>$ or $s_{exp}$/$<$ {\it count rate} $>$ in the case of its upper limit. Magnitudes in UBV Johnson and AB systems can be obtained from the count rate and using the {\it zero point} and count rate--to--flux conversion factors  provided by the {\it Swift} team in their web page.
}\\
\hline\hline
Object (UV Filter)  & $s$[counts/s] & $s_{exp}$[counts/s]& $s/s_{exp}$ &$<${\it count rate}$>$& $s_{frac}$\\
\hline
\endfirsthead
\caption{continued.}\\
\hline\hline
Object (UV Filter)  & $s$[counts/s] & $s_{exp}$[counts/s]& $s/s_{exp}$ &$<${\it count rate}$>$& $s_{frac}$\\
\hline
\endhead
\hline
\endfoot
NQ~Gem (UVM2) &31.2 & 1.1&28.4&204$\pm$8 & 0.15\\
TX CVn (UVW2) & 4.0&1.0&4.0& 241$\pm$2& 0.02\\
TX CVn (UVW2) &  47.7& 1.0& 47.7&  207$\pm$24& 0.23\\
TX CVn (UVM2) &37.0&1.0&37.0&152$\pm$21 &0.24\\
ZZ~CMi (UVM2) & 1.0&0.3 &3.3& 24.8$\pm$0.3& 0.04 \\
ZZ~CMi (UVM2) &3.0&0.3 &10.0 &32.1$\pm$1.1& 0.10 \\
AR Pav (UVM2) & 0.4&0.3&1.3& 38.5$\pm$0.2& 0.01\\
AR Pav (UVM2) & 0.7&0.2&3.5 & 36.3$\pm$0.3& 0.02\\
ER~Del (UVW2) &3.9 & 0.1 & 39.0& 20.1$\pm$1.2& 0.19\\
CD~-27 8661 (UVM2) & 1.5&0.2&7.5 & 17.7$\pm$0.5 & 0.08\\
V627~Cas (UVW2) & 0.05&0.03&1.7& 0.43$\pm$0.03& 0.11\\
V627~Cas (UVW2) & 0.02&0.03&0.67& 0.42$\pm$0.01& $<$ 0.06\\
Hen~3-461  (UVM2) & 0.64 & 0.06 & 10.7&3.8$\pm$0.2& 0.17\\
Wray~16-51 (UVM2) & 0.03&0.03&1.0&0.38$\pm$0.02& 0.09 \\
Wray 16-51 (UVM2) & 0.05&0.03&1.67&0.65$\pm$0.02&0.08 \\
Wray 16-51 (UVM2) & 0.11&0.04&2.75&0.65$\pm$0.06 & 0.17 \\
SY~Mus (UVW2) &4.7&0.4&11.7 & 74.1$\pm$1.6& 0.06\\
CD~-28 3719  (UVW2) &5.4 & 0.2 &27.0 & 37.5$\pm$1.7 & 0.14\\
V443~Her (UVM2) & 5.2&0.4&13.2&85.4$\pm$3.0& 0.06\\
V443~Her (UVM2) & 11.4&0.4&28.5&80.8$\pm$5.1&0.14\\
BD~-21 3873 (UVM2) & 0.4&0.4&1.0&59.8$\pm$0.3&0.01\\
BD~-21 3873 (UVM2) &1.0&0.1&10.0&5.8$\pm$0.4&0.17\\
V748~Cen (UVM2) & 0.4&0.2&2.0& 26.7$\pm$0.2& 0.01\\
V748~Cen (UVM2) & 9.3&0.6&15.5& 141$\pm$3& 0.06\\
UKS~Ce-1 (UVM2) & 0.03&0.03&1.0& 0.14$\pm$0.01& 0.23\\
UKS~Ce-1 (UVM2) & 0.01&0.02&0.6& 0.16$\pm$0.01& $<$ 0.14\\
YY~Her (UVM2) &0.4&0.1&4.0&7.8$\pm$0.2&0.05  \\
CI~Cyg (UVM2) & 0.4&0.3&1.3& 41.4$\pm$0.2& 0.01\\
CI~Cyg (UVM2) & 0.18&0.19&0.97&37.4$\pm$0.1&$<$ 0.01\\
CI~Cyg (UVM2) &1.2&0.2&6.0&36.7$\pm$0.5&0.03\\
FG~Ser (UVM2) & 0.01&0.02&0.5&0.36$\pm$0.01& $<$ 0.06\\
FG~Ser (UVM2) & 0.04&0.03&1.3&0.37$\pm$0.01& 0.11\\
Wray 15-1470 (UVM2) &0.18&0.12&1.5&7.5$\pm$0.1&0.02\\
Wray 15-1470 (UVM2) &0.11&0.20&0.6&7.6$\pm$0.1&$<$ 0.01\\
Wray 15-1470 (UVM2) &0.19&0.10&1.9&7.3$\pm$0.1& 0.03\\
Wray 15-1470 (UVM2) &0.81&0.14&5.8&6.9$\pm$0.5& 0.12\\
Hen 3-863 (UVW2) &  0.03&0.13&0.23&9.83$\pm$0.01&$<$ 0.01 \\
Hen 3-863 (UVW2) &  0.19&0.18&1.05&18.8$\pm$0.1& 0.01 \\
Hen 3-863 (UVW2) &  0.08&0.15&0.5&9.8$\pm$0.1&$<$ 0.01 \\
AS~210 (UVM2) & 0.5&0.1&5.0&10.2$\pm$0.2&0.05\\
StH$\alpha$ 32 (UVM2) &0.1&0.1&1&11.1$\pm$0.1&$<$ 0.01\\
StH$\alpha$ 32 (UUU) &0.3&0.2&1.5&26.9$\pm$0.1&0.01\\
V835~Cen (UVW2)& 0.06&0.08&0.7&2.9$\pm$0.1&$<$ 0.02\\
V835~Cen (UVM2) & 0.03&0.03&1.0&0.82$\pm$0.01&$<$ 0.04\\
BI~Cru (UVW2) & 0.07&0.09&0.8& 5.0$\pm$0.1&$<$ 0.02\\
BI~Cru (UVW2) & 0.10&0.10&1.0 &5.0$\pm$0.1&$<$ 0.02\\
BI~Cru (UVW2) &0.01&0.08 &0.12 &5.1$\pm$0.1&$<$ 0.02\\
BI~Cru (UVW2) &0.07&0.13 &0.54 & 5.3$\pm$0.1&$<$ 0.02\\
AS~289 (UVM2)& 0.04&0.03&1.3&0.46$\pm$0.01&0.08\\
AS~289 (UVM2)& 0.01&0.04&0.25&0.49$\pm$0.01&$<$ 0.08\\
V347~Nor (UVW2)& 0.01&0.02& 0.5&0.12$\pm$0.01&$<$ 0.22\\
V347~Nor (UVW2)& 0.02&0.02& 1.0&0.11$\pm$0.01&$<$ 0.23\\
V347~Nor (UVM2)& 0.01&0.03&0.30&0.11$\pm$0.01&$<$ 0.24\\
V347~Nor (UVM2)& 0.01&0.01&1.0&0.04$\pm$0.01&$<$ 0.26\\
AX~Per (UVW2)&1.13&0.54&2.09&108.7$\pm$0.3&0.01\\
Hen~3-1213 (UVM2)& 0.08&0.08&1.0&3.03$\pm$0.03&0.03\\
LT~Del (UVM2)&0.33&0.15&2.2&10.7$\pm$0.1&0.03\\
Y~Cra (UVM2)&0.28&0.12&2.3 &15.7$\pm$0.1&0.02\\
AS~327 (UVM2)&0.08&0.08&1.0&2.39$\pm$0.05&0.03\\
AS~327 (UVM2)&0.07&0.07&1.0&2.42$\pm$0.03& 0.03\\
AS~327 (UVM2)&0.04&0.06&0.7&2.44$\pm$0.01&$<$ 0.02\\
KX~Tra (UVM2)& 2.4&0.5&4.8&59$\pm$1&0.04\\
KX~Tra (UVM2)& 3.8&0.2&16&57$\pm$1&0.07\\
V366~Car (UVM2)&0.06&0.07&0.8&4.03$\pm$0.02&$<$ 0.02\\
SWIFT~J171951.7-300206 (UUU) & ... & ... & ... & 2.05$\pm$0.14 & ... \\
SWIFT~J171951.7-300206 (UVM2) & ... & ... & ... & $<$0.04 & ... \\
SWIFT~J171951.7-300206 (UVW1) & ... & ... & ... & 0.37$\pm$0.03 & ... \\
SWIFT~J171951.7-300206 (UVW2) & ... & ... & ... & 0.17$\pm$0.02 & ... \\
\label{time}
\end{longtable}
}

\end{document}